\documentclass[a4paper,11pt]{article}

\usepackage{amsmath,amsfonts,amssymb,caption,graphicx}
\usepackage{ushort}
\usepackage{makeidx}
\usepackage{float}
\usepackage{accents}
\usepackage{color}
\usepackage{xcolor}
\usepackage{framed}
\usepackage{versions}
\usepackage{emptypage}
\usepackage{wrapfig}
\usepackage[T1]{fontenc}
\usepackage[utf8]{inputenc}
\usepackage{titlesec}
\usepackage{fancyhdr}
\usepackage{extramarks}
\usepackage[bookmarks]{hyperref}
\usepackage{hyperref}
\usepackage{bbm}

\hypersetup{
    colorlinks=true,   	
    linkcolor=red,      
    citecolor = [rgb]{0 0.7 0},   	
    filecolor=magenta, 	
    urlcolor=blue
}

\newcommand{\Z}{{\rm Z}}

\newcommand{\Xp}{\mbox{\boldmath $X$}}

\newcommand{\Yp}{\mbox{\boldmath $Y$}}

\newcommand{\Zp}{\mbox{\boldmath $Z$}}

\newcommand{\BBP}{{\mathbb P}}

\newcommand{\BBE}{{\mathbb E}}

\def\ba{\begin{align}}
\def\ea{\end{align}}
\def\ban{\begin{align*}}
\def\ean{\end{align*}}

\def\be{\begin{eqnarray}}
\def\ee{\end{eqnarray}}
\def\ben{\begin{eqnarray*}}
\def\een{\end{eqnarray*}}

\def\bqq{\begin{equation}}
\def\eqq{\end{equation}}
\def\bqqn{\begin{equation*}}
\def\eqqn{\end{equation*}}

\def\sq{$\Box$}

\def\qed{\ifmmode\sq\else{\unskip\nobreak\hfil
\penalty50\hskip1em\null\nobreak\hfil\sq
\parfillskip=0pt\finalhyphendemerits=0\endgraf}\fi\par\medbreak}

\newsavebox{\junk}
\savebox{\junk}[1.6mm]{\hbox{$|\!|\!|$}}

\newcommand{\indep}{\perp \!\!\! \perp}

\def\til={{\widetilde =}}

 \def\eq#1/{(\ref{#1})}

\newtheorem{theorem}{Theorem}[section]
\newtheorem{corollary}[theorem]{Corollary}
\newtheorem{proposition}[theorem]{Proposition}

\def\eq#1/{(\ref{e:#1})}

\def\bdes{\begin{description}}
\def\edes{\end{description}}

\def\notes#1{}

\definecolor{mag}{rgb}{0.7,0,0.3}
\definecolor{dgreen}{rgb}{0.1,0.5,0.1}
\definecolor{dred}{rgb}{.8,0,0}
\definecolor{gray}{rgb}{.8,.8,.8}
\definecolor{brown}{rgb}{0.6451,0.3706,0.1745}

\begin{document}

\title{\vspace{-1.1cm}%
Temporally Causal Discovery Tests\\
for Discrete Time Series and Neural Spike Trains}

\author
{
	A. Theocharous
    \thanks{DPMMS,
	University of Cambridge,
	Centre for Mathematical Sciences,
        Wilberforce Road,
	Cambridge CB3 0WB, U.K.
                Email: \texttt{\href{mailto:at771@cam.ac.uk}%
			{at771@.cam.ac.uk}}.
        }
\and
       G.G. Gregoriou
    \thanks{University of Crete, Faculty of Medicine,
	 and Foundation for Research and Technology Hellas, 
	Institute of Applied and Computational
	Mathematics, Heraklion 70013, Greece.
                Email: \texttt{\href{mailto:gregoriou@uoc.gr}%
			{gregoriou@uoc.gr}}.
	G.G.G.\ was supported in part 
	by the Hellenic Foundation for Research and Innovation (H.F.R.I.) 
	under the First Call for H.F.R.I.\ Research Projects to support 
	Faculty members and Researchers and the procurement of high-cost 
	research equipment grant, project number 41.
        }
\and
        P. Sapountzis
    \thanks{Foundation for Research and Technology Hellas, 
	Institute of Applied and Computational
	Mathematics, Heraklion 70013, Greece.
                Email: \texttt{\href{mailto:pasapoyn@iacm.forth.gr}%
			{pasapoyn@iacm.forth.gr}}.
	P.S.\ was supported in part by the Hellenic Foundation 
	for Research and Innovation (H.F.R.I.) 
	under the First Call for H.F.R.I.\ Research Projects to support 
	Faculty members and Researchers and the procurement of high-cost 
	research equipment grant, project number 41.
        }
\and
        I. Kontoyiannis
    \thanks{Statistical Laboratory, DPMMS,
	University of Cambridge,
	Centre for Mathematical Sciences,
        Wilberforce Road,
	Cambridge CB3 0WB, U.K.
                Email: \texttt{\href{mailto:yiannis@maths.cam.ac.uk}%
			{yiannis@maths.cam.ac.uk}}.
	I.K.\ was supported in part by the Hellenic Foundation for Research 
	and Innovation (H.F.R.I.) under the ``First Call for H.F.R.I. Research 
	Projects to support Faculty members and Researchers and the 
	procurement of high-cost research equipment grant,'' project 
	number 1034.
        }
}

\date{\today}

\maketitle

\begin{abstract}
We consider the problem of detecting
causal relationships between discrete 
time series, in the presence of potential confounders. 
A hypothesis test is introduced for 
identifying the temporally causal influence 
of $(x_n)$ on $(y_n)$, 
causally conditioned on a possibly confounding 
third time series $(z_n)$. 
Under natural Markovian modeling assumptions, 
it is shown that the null hypothesis, 
corresponding to the absence of temporally causal 
influence, is equivalent to 
the underlying 
`causal conditional directed information rate'
being equal to zero. The plug-in estimator for
this functional is identified with the 
log-likelihood ratio test statistic for the
desired test. This statistic is shown to be
asymptotically normal under the alternative hypothesis
and asymptotically $\chi^2$ distributed
under the null, facilitating the computation
of $p$-values when used on empirical data.
The effectiveness of the resulting hypothesis test 
is illustrated on simulated data, 
validating the underlying theory.
The test is also employed in the analysis 
of spike train data recorded from neurons 
in the V4 and FEF brain regions of behaving
animals
during a visual attention task. There, the test results
are seen to identify interesting and biologically
relevant information.

\medskip

\end{abstract}

\newpage

\section{Introduction}

\subsection{Information, independence and causality}
\label{s:IIC}

If $\Xp= \{ X_n \}$ and  $\Yp= \{ Y_n \}$ are two finite-valued 
random processes, the \textit{mutual information} $I(X_1^n; Y_1^n)$ 
between $X_1^n=(X_1,X_2,\ldots,X_n)$
and $Y_1^n=(Y_1,Y_2,\ldots,Y_n)$ is,
\bqq
    I(X_1^n; Y_1^n) 
     := H(Y_1^n) - H(Y_1^n| X_1^n)
     = H(Y_1^n) - \sum_{i=1}^n H(Y_i|Y_1^{i-1},X_1^n),
\label{eq:MI}
\eqq
where $H(X)$ denotes the Shannon entropy of a discrete
random variable $X$, 
and with the obvious understanding that $Y_1^0$ is taken to
be a constant, so that $H(Y_1 | Y_1^{0},X_1^n)=H(Y_1|X_1^n)$.
Alternatively,
$$
    I(X_1^n; Y_1^n) 
     = D( P_{X_1^n,Y_1^n} \| P_{X_1^n} P_{Y_1^n} ),
$$
where $D(P\|Q)$ is the relative entropy between
two discrete probability mass functions $P,Q$.

An elementary but key property of
mutual information is that $I(X_1^n;Y_1^n)$ is
always nonnegative and it is zero if and only if
$X_1^n$ and $Y_1^n$ are independent.
Moreover, in communication theory
$I(X_1^n;Y_1^n)$ admits
an operational {\em quantitative} interpretation:
It is the number of bits of
information $X_1^n$ and $Y_1^n$ reveal
about each other. Therefore, $I(X_1^n;Y_1^n)$
is often viewed as a natural,
`universal' measure of dependence between 
random variables.
Partly due to this interpretation,
in recent years mutual information has also been instrumental
in numerous modern applications 
beyond its fundamental role
in classical communication and information 
theory~\cite{cover:book2,gallager:book,csiszar:book2},
in important problems in
statistics~\cite{steuer:02,decampos:06,kinney:14,belghazi:18},
cryptography~\cite{bruss:98,gierlichs:08}
and machine 
learning~\cite{battiti:94,tishby:99,mackay:book,vergara:14,bachman:19},
among other areas.

A closely related but critically different information-theoretic 
functional is the {\em directed information}
$I(X_1^n \to Y_1^n)$ from $X_1^n$ to $Y_1^n$, defined as,
\begin{align}
    I(X_1^n \to Y_1^n) 
	&:= H(Y_1^n) - \sum_{i=1}^n H(Y_i|Y_1^{i-1},X_1^i)
	\label{eq:DI}\\
	&= \sum_{i=1}^n I(X_1^i ; Y_i | Y_1^{i-1}).
	\label{eq:DI2}
\end{align}
A comparison of the expressions in~(\ref{eq:MI}) and~(\ref{eq:DI})
gives a first indication of the `causal' nature of directed
information, one of the main themes of this work.

The corresponding key property of directed information, 
which is evident from~(\ref{eq:DI2}), is that
$I(X_1^n\to Y_1^n)$ is always nonnegative
and it is zero if and only if $Y_i$ is conditionally
independent of $X_1^i$, given its past $Y_1^{i-1}$,
for each $i$. Therefore, directed information can 
naturally be viewed as a measure of the degree to 
which $X_1^n$ influences $Y_1^n$
in a {\em temporally causal manner}.
In fact, in its original communication-theoretic 
context~\cite{marko:73,massey:90}, directed information 
was given a corresponding quantitative interpretation as the amount 
of information about the `channel input' $X_1^n$ 
contained in the `channel output' $Y_1^n$, 
when at each stage $i$ the channel encoder 
sending $X_i$ is given access to `causal feedback',
that is, to the earlier channel output $Y_1^{i-1}$;
see, e.g.,~\cite{kramer:phd,kim:08,permuter-et-al:09}. 

As a measure of temporally causal dependence, 
directed information has found applications in
numerous areas including
distributed hypothesis testing~\cite{gunduz-erkip:07},
portfolio theory~\cite{weissman-et-al:11},
network communications and control~\cite{kramer:14}, 
causal estimation~\cite{weissman-et-al:13},
dynamic networks~\cite{haskovec:15},
sensor networks~\cite{rahimian:13} and,
most relevant to the development in this paper,
in temporal causality testing~\cite{skoularidou-K:16}.

The present work focuses on a 
third information-theoretic functional,
the {\em causal conditional directed information}
(CCDI)
from $X_1^n$ to $Y_1^n$ causally conditioned on $Z_1^n$, defined as,
\begin{align}
I(X_1^n \to Y_1^n \| Z_1^n) 
&:= 
	H(Y_1^n\| Z_1^n) -H(Y_1^n\| X_1^n, Z_1^n) 
	\label{eq_caus_cond_dir_inf}\\
&:=  
	\sum_{i=1}^n H(Y_i|Y_1^{i-1},Z_1^i) 
	- \sum_{i=1}^n H(Y_i|Y_1^{i-1},X_1^i, Z_1^i)
	\label{eq:CCDI}\\
&=
	\sum_{i=1}^n I(X_1^i;Y_i|Y_1^{i-1},Z_1^i).
	\nonumber
\end{align}
Here, $I(X_1^n \to Y_1^n\| Z_1^n)$ is nonnegative 
and it is equal to zero if and only if, for each $i$,
$Y_i$ is conditionally independent of $X_1^i$ given 
its past $Y_1^{i-1}$ and the past and present values of the 
third process $Z_1^i$:
\begin{align}
&[Z_1,Z_2,\ldots,Z_{i-1},Z_i]\nonumber\\
&X_1,X_2,\ldots,X_{i-1},X_i\label{eq:graphtest}\\
&[Y_1,\,Y_2,\ldots,\,Y_{i-1}],Y_i\nonumber
\end{align}
So $I(X_1^n \to Y_1^n\| Z_1^n)$ is zero iff
the $\Xp$ process up to time $i$ has no 
influence on $Y_i$, given the past $Y_1^{i-1}$
of the $\Yp$ process and the values $Z_1^i$ 
up to time $i$ of
a potentially confounding process $\Zp$.
In other words, if our goal is to predict
$Y_i$, then
$I(X_1^n \to Y_1^n\| Z_1^n)=0$ implies that,
if we already know its past
$Y_1^{i-1}$ and also $Z_1^i$, knowing $X_1^i$
has no additional predictive value.
Therefore, and in view of the previous 
discussion, $I(X_1^n\to Y_1^n\|Z_1^n)$
can be interpreted
as a measure of the temporally causal
influence of $\Xp$ on $\Yp$, in the
presence of possible confounding in the
form of $\Zp$.
Indeed, after its introduction 
by Kramer~\cite{kramer:phd} in the context
of channel coding, 
CCDI has found
applications along these lines
in numerous areas including
causality detection~\cite{shibuya:11},
pattern analysis~\cite{haruna:13},
time series prediction~\cite{cabuz:18},
model selection~\cite{gao:17},
neuroscience~\cite{neurspik}, 
and finance~\cite{schamberg:20}.

\subsection{Temporal causality with confounders}

Two closely related problems are treated in this work.
The first is the identification of effective estimators
for the CCDI from empirical data,
along with the analysis of their performance. 
The second is the 
development of an efficient hypothesis test 
for temporally causal dependence in time
series, when a potentially confounding time series
is also present.
Our results along both of these directions
can be seen as generalizations of the corresponding 
results in~\cite{skoularidou-K:16}, with the important
difference that a third, potentially confounding 
process, $\Zp$, is now included.

Indeed, in observational studies such as in the present
context of time series analysis, allowing for potential confounders 
is extremely important, as it often facilitates the distinction 
between simple correlation and actual causality~\cite{greenland:99}.

\medskip

\noindent
{\bf CCDI rate estimation. }
We begin, in Section~\ref{s:CCDI}, by identifying
natural Markovian conditions under which, 
the \textit{CCDI rate} 
from $\Xp$ to $\Yp$ causally conditioned on  $\Zp$,
\begin{equation}
\label{eq_3}
    I(\Xp \to \Yp \| \Zp) 
:= \lim_{n \to \infty} \frac{1}{n} I(X_1^n \to Y_1^n \| Z_1^n),
\end{equation}
exists and can be expressed as 
a finite-dimensional
conditional mutual information.

In Section~\ref{sec_plug_in_estor_and_thms} we 
consider the following simple estimator for
$I(\Xp \to \Yp \| \Zp)$. Utilizing the fact that
our expression for the CCDI rate
is as a functional of a collection
of $(k + 1)$-dimensional distributions,
we define the plug-in 
estimator $\hat{I}_n^{(k)}(\Xp \to \Yp \| \Zp )$ 
as the same functional of the corresponding
empirical distributions induced by the data.
The parameter $k$ here denotes maximal length 
of the Markov memory order assumed to be present
in the data, and $n$ is the data length.
When the joint process $(\Xp,\Yp,\Zp)$ is ergodic
we can immediately conclude that the 
the estimator $\hat{I}_n^{(k)}(\Xp \to \Yp \| \Zp )$ 
is asymptotically 
consistent with probability~1. 

Our main theoretical results give precise
descriptions
of the finer asymptotic behavior of 
$\hat{I}_n^{(k)}(\Xp \to \Yp \| \Zp )$.
In Theorem~\ref{thmN} we show that,
if the CCDI rate $I(\Xp \to \Yp \| \Zp)$ is strictly
positive, then, under appropriate conditions, 
the plug-in estimator is asymptotically
approximately Gaussian:
\begin{equation*}
   \hat{I}_n^{(k)}(\Xp \to \Yp \| \Zp ) 
   \approx N \Big(  I(\Xp \to \Yp \| \Zp) , \frac{\sigma^2}{n} \Big),
	\quad\mbox{for large}\;n.
\end{equation*}
The variance $\sigma^2$ is also identified in Theorem~\ref{thmN}. 
Therefore, the plug-in estimator converges at a rate $O(1/\sqrt{n})$ 
in probability when 
$I(\Xp \to \Yp \| \Zp)>0$.

In Theorem~\ref{thmX} we show that,
if $I(\Xp \to \Yp \| \Zp)$ is zero, then,
again under appropriate conditions,
the plug-in estimator is asymptotically
$\chi^2$ distributed:
\begin{equation}
   2n\hat{I}_n^{(k)}(\Xp \to \Yp \| \Zp ) \approx 
	\chi^2\Big(\ell^k t^{k+1} (m^{k+1}-1) (\ell-1)\Big),
	\quad\mbox{for large}\;n.
   \label{eq_5d} 
\end{equation}
Here $m, \ell, t$ are the alphabet sizes of 
$\Xp$, $\Yp$, $\Zp$, respectively.
In this case, $ \hat{I}_n^{(k)}(\Xp \to \Yp \| \Zp )$ 
converges at the faster rate $O(1/n)$ 
in probability.
Importantly, its limiting distribution only depends on
the known parameters $m, \ell, t,k$. 
Also, in Corollary~\ref{cor_3.4} we show that
$\hat{I}_n^{(k)}(\Xp \to \Yp \| \Zp )$
is asymptotically optimal in that it achieves
the fastest possible $L^1$ convergence rate.

\medskip

\noindent
{\bf Temporal causality testing. }
In Section~\ref{sec_hypothesis_test} we describe how the
asymptotic result~(\ref{eq_5d})
can be used to develop a hypothesis test
for the temporally causal influence of $\Xp$ 
on $\Yp$, in the presence of a potential confounder $\Zp$. 

First we recall that,
under appropriate assumptions, the CCDI rate 
$I(\Xp \to \Yp \| \Zp )$ can be expressed as 
a conditional mutual information. This
conditional mutual information is zero 
exactly under the null hypothesis of the
desired test, that is, if and only if
$Y_{k+1}$ is conditionally
independent of $X_1^{k+1}$, given the past
values $Y_1^k$ and the values $Z_1^{k+1}$ of the 
potentially confounding process $\Zp$ up to time $k+1$;
cf.~(\ref{eq:graphtest}).
Then, we show in Proposition \ref{prop_3.5} 
that the scaled version of the plug-in estimator
as in~(\ref{eq_5d}), namely,
$\Delta_n:=2n\hat{I}_n^{(k)}(\Xp \to \Yp \| \Zp )$,
is exactly equal to the log-likelihood ratio
test statistic for this null hypothesis.
Finally we observe that, since the limiting
distribution of $\Delta_n$
under the null 
is $\chi^2$ with a known number of degrees
of freedom~(\ref{eq_5d}), it can be used as a classical
statistic to provide a $p$-value for
the desired hypothesis test.

The performance of this test on
simulated data is examined in
Section~\ref{s:simulations}.
When no potential
confounding process $\Zp$ is present,
our results as well as the corresponding
hypothesis test reduce to those
developed in~\cite{skoularidou-K:16}.
Since no experimental results were 
presented there, we we also examine 
the performance of this simpler
test in Section~\ref{s:simulations}.
Empirical results on synthetic
data are shown to validate 
our theoretical results and 
illustrate the effectiveness 
of both tests. 

In Section~\ref{s:spikes} we apply these tests to a large
data set consisting of neural spike trains recorded from 
the V4 and FEF brain regions of behaving non-human primates 
during a behavioral task in which attention was directed selectively 
to a specific visual stimulus among other distracting stimuli.
The results obtained
indicate that there are often temporally causal influences
in both directions between pairs of neurons in the
two different regions. Moreover, these influences
are confounded by the experimental stimuli, which 
appear to affect multiple neurons in both regions
simultaneously.
The code used for the 
hypothesis tests carried out in
Sections~\ref{s:simulations} and~\ref{s:spikes}
is available online at:
\href{https://github.com/Andreas947/github_temporal_causality}%
{\texttt{github.com/Andreas947/github\_temporal\_causality}}.

Finally, the Appendix contains the proofs of all our theoretical
results.

\medskip

\noindent \textbf{Transfer entropy.} 
A functional very closely related to directed information 
is the \textit{transfer entropy},
introduced by Schreiber in 2000~\cite{schreiber:00}, 
and developed further in numerous directions,
including the introduction of the {\em partial transfer entropy}
corresponding to the causal conditional directed information; see,
e.g.,~\cite{amblard:12,faes:13,kugiumtzis:13}.
The estimation of transfer entropy from empirical data
via maximum-likelihood-like methods is explored 
in~\cite{barnett:12,anderson:14},
where the main emphasis appears to be 
less on obtaining rigorous limit theorems and 
more on identifying the potential asymptotic behaviour 
of the relevant estimators.

Although, formally,
these two functionals are near-identical, 
we note that in the present context -- where we seek to 
provide useful and meaningful interpretations of the
results obtained -- directed information,
the associated CCDI rate,
and its extensions, are accompanied by operational
characterizations stemming from ``coding''-like theorems
in various communications scenarios and statistical
problems, as indicated in Section~\ref{s:IIC}.
These considerations also inform the other question 
examined in this work, namely the 
estimation of the directed information, viewed as
a related but separate problem from that of 
establishing the presence or absence of
temporally causal influence.

\medskip

\noindent \textbf{Granger causality.} 
Historically, most temporal causality tests
for time series, their applications,
and their connections with
conditional independence testing,
have their origin in the notion of
{\em Granger causality} in econometrics.
Granger's original work~\cite{granger:69}
was based on a linear autoregressive model,
within which the same family of 
conditional independence hypotheses 
as those described in Section~\ref{s:IIC} above
were tested: A process $\Xp$ is said
to have a temporally causal influence
on $\Yp$ whenever the conditional independence
hypothesis ``$Y_n\indep X_1^n\mid Y_1^{n-1}$''
can be rejected. Tests based on the same general premise
have been since developed for generalized linear models
as well as many other model classes;
see~\cite{shojaie:22} for a recent review of 
extensions and applications. In particular,
this connection
was emphasized for the case of discrete-valued
processes by 
Geweke~\cite{geweke:84} and
for Gaussian models by 
Barnett, Barrett and Seth~\cite{barnett:09},
who also noted that the transfer entropy can
be used as statistic for this test.
The corresponding (and essentially equivalent)
connection between Granger causality and directed 
information 
is discussed extensively in~\cite{amblard:12}.
Work on the theoretical analysis and estimation 
of the CCDI includes~\cite{quinn:15}, where directed 
information graphs are used to represent networks 
of stochastic processes. 

\medskip

\noindent
{\bf Temporally causal influence. }
As the terminology describing causal relationships is
quite varied, intuitively appealing, and easy to misinterpret, 
we wish to explicitly clarify the usage adopted in this
work. We say that a process $\Xp$ has
a temporally causal influence of $\Yp$ if
$Y_n$ is {\em not} conditionally independent of $X_1^n$,
given the past $Y_1^{n-1}$. Of course this
is quite different from 
saying that the $\Xp$ process has some sort of 
deterministic or ``directed'' influence on $\Yp$,
and it does not preclude the case where in fact it is $\Yp$
that somehow influences $\Xp$.
Rather, it stems from what has become standard
usage in statistical time series analysis.
Granger in his original 1969
work~\cite{granger:69} defined statistical causality
in terms of {\em incremental predictability}, 
namely,
in terms of whether the forecasts of the future values 
of $\Yp$ can be improved if -- besides the past values of $\Yp$
and all other available information -- the current
and lagged values of $\Xp$ are also taken into account.
Since then, this point of view along with its numerous
instantiations within different model classes has become
a standard part of statistical nomenclature.
Importantly, Granger's perspective is indeed 
formally equivalent to the conditional independence
test underlying the general approach taken in this work.
The original development of these ideas began 
with~\cite{granger:69,granger:74}, and a broader
discussion of causality and related terminology
in the context of time series
can be found, e.g., in the text~\cite{kirchgassner:book}.

\newpage

\section{Causal conditional directed information}

\subsection{Basic definitions}

Let $X$ be a discrete random variable with values in a finite set $A$
and probability mass function $P_X(x)=\BBP(X=x)$, $x \in A$. 
The entropy of $X$ is
$H(X) =H(P_X) = -\sum_{x \in A} P_X(x) \log P_X(x)$,
where `$\log$' denotes the natural logarithm throughout the paper. 
The conditional entropy $H(X|Y)$ of $X$ given $Y$ is 
$H(X|Y)= H(X,Y)-H(Y)$,
the mutual information is defined as in~(\ref{eq:MI}),
and the conditional mutual information between 
$X$ and $Y$ given $Z$ is,
\begin{equation*}
    I(X; Y | Z) = H(X|Z) + H(Y|Z)- H(X,Y|Z).
\end{equation*}
The relative entropy 
between two probability mass functions $P,Q$
on the same alphabet $A$ is,
$D(P\|Q)= \sum_{x \in A} P(x) \log [P(x)/Q(x)]$.
We write $X_i^j$ for a vector of random variables 
$(X_i, X_{i+1},\ldots  , X_j)$, $i \leq j$, 
and similarly $a_i^j=(a_i, a_{i+1}, \ldots , a_j) \in A^{j-i+1}$, 
for a string of symbols from a finite set $A$. 
The causal conditional entropy  $H(Y_1^n \| Z_1^n)$ between
two blocks of random variables $Y_1^n$ and $Z_1^n$ is defined
as,
\begin{equation*}
    H(Y_1^n \| Z_1^n)=\sum_{i=1}^n H(Y_i|Y_1^{i-1},Z_1^i),
\end{equation*}
and the corresponding 
causal conditional entropy rate between
the processes $\Yp=\{Y_n\}$ and $\Zp=\{Z_n\}$ is,
$$H(\Yp\|\Zp)=\lim_{n\to\infty}\frac{1}{n}
    H(Y_1^n \| Z_1^n),
$$
whenever the limit exists.

\subsection{The CCDI rate of Markov chains}
\label{s:CCDI}

Suppose $\Xp= \{ X_n \}$, $\Yp= \{ X_n \}$ and $\Zp= \{ Z_n \}$ are three arbitrary random processes, with values in the finite 
alphabets $A$, $B$ and $C$ respectively. 
Recall the definitions of the causal conditional directed
information (CCDI)
and the CCDI rate from $\Xp$ to $\Yp$ causally conditioned on $\Zp$ 
in~(\ref{eq_caus_cond_dir_inf})
and~(\ref{eq_3}), respectively.

Our first result provides conditions under which 
the CCDI rate exists and can be expressed in a finite-dimensional
form; its proof is given in Appendix~\ref{app:proposition}.

\begin{proposition}[CCDI rate]
\label{prop_irr,aper}
Suppose $\{(X_n, Y_n, Z_n)\}$ is an irreducible and aperiodic Markov 
chain of order no larger than $k \geq 1$.
Let $(X_{-k+1}^0,Y_{-k+1}^0,Z_{-k+1}^0)$ have
an arbitrary initial distribution,
and write
$\{(\Bar{X}_n, \Bar{Y}_n, \Bar{Z}_n)\}$ for the stationary 
version of $\{(X_n, Y_n, Z_n)\}$. Then: 
    \begin{enumerate}
        \item[$(i)$] The causal conditional 
	entropy rate $H(\Yp \| \Zp )$ exists and it equals,
        \begin{equation}
            H(\Yp \| \Zp )
	    = \lim_{n \to \infty} \frac{1}{n} H(\Bar{Y}_1^n \| \Bar{Z}_1^n).
		\label{eq:ccH}
        \end{equation}
        \item[$(ii)$] The CCDI rate 
	$I(\Xp \to \Yp \| \Zp )$ exists and it equals,
        \begin{equation*}
            I(\Xp \to \Yp \| \Zp )
	= H(\Yp \| \Zp ) 
	-H(\Bar{Y}_0 | \Bar{Y}_{-k}^{-1},\Bar{X}_{-k}^0, \Bar{Z}_{-k}^0).
        \end{equation*}
        \item[$(iii)$] If $ \{(Y_n, Z_n) \}$ is also a Markov chain of order no larger than k, then $I(\Xp \to \Yp \| \Zp )$ further simplifies to,
        \begin{equation*}
            I(\Xp \to \Yp \| \Zp ) = I(\Bar{Y}_0 ; \Bar{X}_{-k}^0 | \Bar{Y}_{-k}^{-1}, \Bar{Z}_{-k}^0).
        \end{equation*}
    \end{enumerate}
\end{proposition}

Before proceeding with our main theoretical results,
some remarks are in order.
First, we note that the CCDI rate 
admits important operational interpretations. 
If $\{(X_n, Y_n, Z_n)\}$ and $\{( Y_n, Z_n)\}$ 
are both stationary $k$th order Markov chains 
then, by the data processing property of conditional 
mutual information~\cite{cover:book2} 
and Proposition~\ref{prop_irr,aper}~$(iii)$,
we have,
        \begin{equation*}
            I(\Xp \to \Yp \| \Zp ) = I(Y_0 ; X_{-k}^0 | Y_{-k}^{-1}, Z_{-k}^0) = I(Y_0 ; X_{-\infty}^0 | Y_{-\infty}^{-1}, Z_{-\infty}^0).
        \end{equation*}
Therefore, the CCDI rate in this case is zero if and only if 
each $Y_i$ is conditionally independent of the past
$X_{-\infty}^i$ of the $\Xp$ process,
given its own past $Y_{-\infty}^{i-1}$
and the values $Z_{-\infty}^i$ up to time $i$ of the potentially confounding
process $\Zp$. This validates our intuition that the CCDI 
rate is only zero in the absence of causal conditional influence,
and it also naturally leads to the problem of estimating
the CCDI rate.

In fact, we argue that an estimate
of $I(\Bar{Y}_0 ; \Bar{X}_{-k}^0 | \Bar{Y}_{-k}^{-1}, \Bar{Z}_{-k}^0)$
is a useful statistic even if it is not equal
to the CCDI rate $I(\Xp \to \Yp \| \Zp )$. 
Consider a general stationary Markov chain $\{(X_n, Y_n, Z_n)\}$,
but without any further assumptions on $\{( Y_n, Z_n)\}$,
so that,
    \begin{equation*}
            I(Y_0 ; X_{-k}^0 | Y_{-k}^{-1}, Z_{-k}^0)= I(Y_0 ; X_{-\infty}^0 | Y_{-k}^{-1}, Z_{-k}^0) \geq I(Y_0 ; X_{-\infty}^0 | Y_{-\infty}^{-1}, Z_{-\infty}^0).
    \end{equation*}
Here, the quantity $I(Y_0 ; X_{-k}^0 | Y_{-k}^{-1}, Z_{-k}^0)$ is not 
equal to the CCDI rate $I(\Xp \to \Yp \| \Zp )$, but being able to 
estimate its value may still offer useful evidence for the 
temporally causal influence of $\Xp$ on $\Yp$ causally conditioned
on $\Zp$. 
For example, knowing that $I(Y_0 ; X_{-k}^0 | Y_{-k}^{-1}, Z_{-k}^0)$ 
is ``significantly far'' from zero, would still provide strong
evidence for the fact that $\Xp$ has 
a temporally causal influence on $\Yp$ causally conditioned on $\Zp$.
    
\subsection{CCDI rate estimation}
\label{sec_plug_in_estor_and_thms}

Let $(X_{-k+1}^n, Y_{-k+1}^n, Z_{-k+1}^n)$ be a random
sample from the joint process $\{(X_n, Y_n, Z_n)\}$.
The joint empirical distribution induced by this sample
on $A^{k+1} \times B^{k+1} \times C^{k+1}$ is,
\begin{equation}
\label{eq_9}
    \hat{P}_{X_{-k}^0,Y_{-k}^0,Z_{-k}^0,n}(a_0^k, b_0^k, c_0^k)= \frac{1}{n} \sum_{i=1}^n \mathbb{I}_{ \{ X_{i-k}^i=a_0^k,Y_{i-k}^i=b_0^k,Z_{i-k}^i=c_0^k \} },
\end{equation}
for all $a_0^k \in A^{k+1}$, $b_0^k \in B^{k+1}$ and $c_0^k \in C^{k+1}$.
In view of Proposition~\ref{prop_irr,aper},
we can then define
the \textit{plug-in estimator} for the 
CCDI rate $I(\Xp \to \Yp \| \Zp )$ by:
\begin{equation}
\label{eq_10}
    \hat{I}_n^{(k)}(\Xp \to \Yp \| \Zp ) = I(\hat{Y}_0 ; \hat{X}_{-k}^0 | \hat{Y}_{-k}^{-1}, \hat{Z}_{-k}^0), \quad \text{where }  (X_{-k}^0,Y_{-k}^0,Z_{-k}^0) \sim \hat{P}_{X_{-k}^0,Y_{-k}^0,Z_{-k}^0,n}.
\end{equation}

If the $(k+1)$-dimensional chain 
$\{( X_{n-k}^n, Y_{n-k}^n, Z_{n-k}^n)\}$ 
is irreducible and aperiodic, then the
ergodic theorem for Markov chains~\cite{chung:book} implies that 
the empirical distributions $\hat{P}_{X_{-k}^0,Y_{-k}^0,Z_{-k}^0,n}$ 
converge almost surely (a.s.) to the true underlying marginals
$P_{\Bar{X}_{-k}^0,\Bar{Y}_{-k}^0,\Bar{Z}_{-k}^0}$, as $n \to \infty$,
and therefore,
the plug-in estimator $\hat{I}_n^{(k)}(\Xp \to \Yp \| \Zp )$ 
also converges a.s.\ to the desired 
value $I(\Bar{Y}_0;\Bar{X}_{-k}^0|\Bar{Y}_{-k}^{-1}, \Bar{Z}_{-k}^0)$. 
Its finer asymptotic behavior is described in our two main 
theoretical results, stated next.

\begin{theorem}[Asymptotic normality]
\label{thmN}
Suppose $\{(X_n, Y_n, Z_n)\}$ is an irreducible and
aperiodic Markov chain of order no larger than $k \geq 1$,
with values in the finite alphabet $A\times B\times C$
and with an arbitrary initial distribution. 
Assume that the bivariate 
process $\{(Y_n, Z_n)\}$ is also a Markov chain of order 
no larger than $k \geq 1$.

If $\Xp$ does not have a temporally causal influence 
on $\Yp$ causally conditioned on $\Zp$, equivalently, 
if the CCDI rate $I(\Xp \to \Yp \| \Zp )>0$, then,
        \begin{equation}
            \sqrt{n}\Big[ \hat{I}_n^{(k)}(\Xp \to \Yp \| \Zp )
	 	- I(\Xp \to \Yp \| \Zp )\Big]  
		\xrightarrow[]{\mathcal{D}} N (0, \sigma^2), \quad \text{as } 
		n \to \infty,
        \label{eq_normal}
        \end{equation}
where the variance $\sigma^2$ is given by the following limit, 
which exists and is finite:
$$
\sigma^2
= \lim_{n \to \infty} \frac{1}{n} \textup{Var} 
\left\{   \sum_{i=1}^n  \log 
\left( \frac{ P_{\Bar{X}_{-k}^0,\Bar{Y}_0|\Bar{Y}_{-k}^{-1},\Bar{Z}_{-k}^0} 
(X_{i-k}^i, Y_i | Y_{i-k}^{i-1}, Z_{i-k}^i)}{P_{\Bar{Y}_0|\Bar{Y}_{-k}^{-1},
\Bar{Z}_{-k}^0} (Y_i|Y_{i-k}^{i-1}, Z_{i-k}^i) 
P_{\Bar{X}_{-k}^0|\Bar{Y}_{-k}^{-1},
\Bar{Z}_{-k}^0} (X_{i-k}^i |Y_{i-k}^{i-1}, Z_{i-k}^i)} \right)  \right\}.
$$
\end{theorem}
        
Theorem~\ref{thmN} is proved in Appendix~\ref{proofthmN},
where a slightly stronger result is in fact established:
Without the assumption 
that the bivariate 
process $\{(Y_n, Z_n)\}$ is also a Markov chain,
we prove~(\ref{eq_normal}) with
$I(\Bar{Y}_0;\Bar{X}_{-k}^0|\Bar{Y}_{-k}^{-1},\Bar{Z}_{-k}^0)$
in place of $I(\Xp \to \Yp \| \Zp )$.
The result of the theorem then follows from this
in combination with Proposition~\ref{prop_irr,aper}~$(iii)$.

An analogously more general version of Theorem~\ref{thmX}
is proved in Appendix~\ref{proofthmX}.

\begin{theorem}[$\chi^2$ convergence]
\label{thmX}
Let $\{(X_n, Y_n, Z_n)\}$ be a Markov chain of order no larger 
than $k \geq 1$
with an all positive transition matrix Q on the finite 
alphabet $A \times B \times C$, 
\begin{align*}
Q(a_k, 
& 
	b_k,c_k|  a_0^{k-1}, b_0^{k-1}, c_0^{k-1})\\ 
&= 
	\BBP(X_n= a_k, Y_n = b_k, Z_n = c_k| X_{n-k}^{n-1}=a_0^{k-1},
	 Y_{n-k}^{n-1}=b_0^{k-1}  , Z_{n-k}^{n-1}=c_0^{k-1})> 0,
\end{align*}
for all $a_0^k \in A^{k+1}$, $b_0^k \in B^{k+1}$ and $c_0^k \in C^{k+1}$,
and with an arbitrary initial distribution.
Assume that the bivariate 
process $\{(Y_n, Z_n)\}$ is also a Markov chain of order 
no larger than $k \geq 1$.

        If $I(\Xp\to\Yp\|\Zp) =0$,
then,
\begin{equation}
\label{eq_12}
    2n \hat{I}_n^{(k)}(\Xp \to \Yp \| \Zp )  \xrightarrow[]{\mathcal{D}} \chi^2\Big(\ell^k t^{k+1} (m^{k+1}-1) (\ell-1)\Big)   , \quad \text{as } n \to \infty,
\end{equation}
where $m = |A|$, $\ell = |B|$ and $t= |C|$ are the sizes of the alphabets $A,B$ and $C$ respectively.
\end{theorem}

The most important observation from the results of 
Theorems~\ref{thmN} and~\ref{thmX} is the dichotomy in 
the rate of convergence in the cases of presence 
and absence of causal conditional influence. 
If $\Xp$ does {\em not} have a temporally causal influence 
on $\Yp$ causally conditioned $\Zp$ or, equivalently, 
if $I(\Xp \to \Yp \| \Zp )=0$, 
then the plug-in estimator $\hat{I}_n^{(k)}(\Xp \to \Yp \| \Zp )$ 
converges to zero at a rate $O(1/n)$. On the other hand, 
if such causal conditional influence is present, then the 
plug-in estimator $\hat{I}_n^{(k)}(\Xp \to \Yp \| \Zp )$ 
converges 
to the CCDI rate $I(\Xp \to \Yp \| \Zp )$ 
at the slower rate of $O(1/\sqrt{n})$.

Note that
Theorem~\ref{thmX} is stated under the assumption
that all the elements of the transition matrix of 
the joint process $\{(X_n, Y_n, Z_n)\}$ 
are nonzero. This is strictly stronger than
the irreducibility and aperiodicity assumption
in  Theorem~\ref{thmN}. Although we will not prove
this here, the positivity assumption 
can be significantly relaxed; for 
example,~\cite[Theorem~5.2]{billingsley:markov} gives weaker 
conditions under which the same conclusions can be obtained, 
e.g., by considering classes of ergodic chains whose transition 
matrices are allowed to contain zero probabilities at the same 
state transitions.

As it turns out, the proof of Theorem~\ref{thmN}
can be extended to provide bounds on the a.s.\ and
$L^1$ convergence rate of the estimator
$\hat{I}_n^{(k)}(\Xp \to \Yp \| \Zp )$. These
are stated in the corollary below, and proved in 
Appendix~\ref{cor_3.4}. In fact, in view of the
discussion and the earlier 
results in~\cite{weissman-et-al:di,skoularidou-K:16},
the $L^1$ rate of convergence of the plug-in
estimator is optimal among all $L^1$-consistent
estimators of the CCDI rate.
   
\begin{corollary}
\label{cor_3.4}
Let $\{(X_n, Y_n, Z_n)\}$ be an irreducible and aperiodic Markov chain 
of order no larger than $k \geq 1$, with an arbitrary initial distribution. 
As $n \to \infty$, the plug-in estimator satisfies:
\begin{align}
  \hat{I}_n^{(k)}(\Xp \to \Yp \| \Zp )- I(\Bar{Y}_0;\Bar{X}_{-k}^0|\Bar{Y}_{-k}^{-1},\Bar{Z}_{-k}^0) &= O \left( \sqrt{\frac{\log \log n}{n}} \right)  
	\quad a.s.,          \label{eq_13}
  \\  \BBE \left[ \left| \hat{I}_n^{(k)}(\Xp \to \Yp \| \Zp )- I(\Bar{Y}_0;\Bar{X}_{-k}^0|\Bar{Y}_{-k}^{-1},\Bar{Z}_{-k}^0)  \right| \right] &= O \left( \frac{1}{\sqrt{n}} \right).    \label{eq_14}
\end{align}
\end{corollary}

\noindent
{\bf Efficiency. }
In addition to achieving the optimal 
$O(1/\sqrt{n})$ rate in $L^1$, the CCDI estimator 
$\hat{I}_n^{(k)}(\Xp \to \Yp \| \Zp )$ is in
fact {\em efficient}, in that its asymptotic
variance in~(\ref{eq_normal}) of Theorem~\ref{thmN} 
is minimal. To see that, first we recall that
the maximum likelihood estimates (MLEs) 
$\hat{P}_{X_{-k}^0,Y_{-k^0},Z_{-k^0},n}$ of the
parameters of the chain $\{(X_n,Y_n,Z_n)\}$
are themselves efficient~\cite{greenwood:95};
see the proof of Proposition~\ref{prop_3.5}
for the fact that these $\hat{P}$ are indeed MLEs.
And since the MLE is functionally equivariant,
cf.~\cite{lehmann:book,olive:04}, the estimator
$\hat{I}_n^{(k)}(\Xp \to \Yp \| \Zp )$ is also
efficient, as a function of the MLEs $\hat{P}$.

\subsection{Testing for causal conditional influence}
\label{sec_hypothesis_test}

Consider the problem of testing whether or not the 
time series samples generated by the process
$\Xp$ have a temporally causal influence on $\Yp$,  
causally conditioned on the samples $\Zp$. In the present context, this 
corresponds to to testing the null hypothesis that 
each random variable $Y_i$ 
is conditionally independent of $X_{i-k}^i$ given $Y_{i-k}^{i-1}$ 
and $Z_{i-k}^i$, within the larger hypothesis that the 
joint process 
$\{(X_n, Y_n, Z_n)\}$ is a $k$th order Markov chain on 
$A \times B \times C$, with all positive transitions. 
Without loss of generality, we take the alphabets of 
$\Xp, \Yp$ and $\Zp$ to be 
$A = \{1, 2,\ldots ,m\}$,
 $B = \{1, 2, \ldots ,\ell \}$ 
and $C = \{1, 2, \ldots,t\}$ respectively.

Formally,
each positive transition matrix $Q=Q_\theta$ 
for the process $\{(X_n,Y_n,Z_n)\}$ can be indexed 
by a parameter vector $\theta$ taking values in 
an $m^k \ell^k t^k (m \ell t-1)$-dimensional
open set $\Theta$; 
see the proof of Theorem~\ref{thmX} 
in Appendix~\ref{proofthmX} for explicit details.
The null hypothesis corresponding to each $Y_i$ 
being conditionally independent of $X_{i-k}^i$ 
given $Y_{i-k}^{i-1}$ and $Z_{i-k}^i$, 
is described by transition matrices $Q_\theta$
that can be expressed as,
\begin{equation}
    \label{eq_16}
Q_\theta(a_0,b_0,c_0|a_{-k}^{-1},b_{-k}^{-1},c_{-k}^{-1}) 
= Q_\theta^{x,z}(a_0,c_0|a_{-k}^{-1},b_{-k}^{-1},c_{-k}^{-1})
Q_\theta^y (b_0|b_{-k}^{-1},c_{-k}^0).
\end{equation}
This collection of transition matrices can
be indexed by parameters in a lower-dimensional parameter 
set $\Phi$ that can be embedded in $\Theta$ 
via a map $h: \Phi \to \Theta$, 
with the property that all $Q_{h(\phi)}$ satisfy the 
conditional independence property in~\eqref{eq_16}.

We employ a classical likelihood ratio for testing
the null hypothesis $\Phi$ within the general model 
$\Theta$. The log-likelihood 
$L_n(X_{-k+1}^n,Y_{-k+1}^n,Z_{-k+1}^n;\theta)$ 
of the sample $(X_{-k+1}^n,Y_{-k+1}^n,Z_{-k+1}^n)$ under the 
distribution corresponding to $\theta$ can be expressed as,
\begin{align}
        L_n(X_{-k+1}^n,Y_{-k+1}^n,Z_{-k+1}^n;\theta) 
&=  \log[\BBP_\theta(X_1^n,Y_1^n,Z_1^n | X_{-k+1}^0,Y_{-k+1}^0,Z_{-k+1}^0 )] 
	\nonumber\\ 
&= \log \left( \prod_{i=1}^n Q_\theta(X_i,Y_i,Z_i | X_{i-k}^{i-1},Y_{i-k}^{i-1},Z_{i-k}^{i-1} ) \right) ,    \label{eq_17} 
\end{align}
where $\BBP_\theta$ denotes the law of the 
$k$th order Markov process $\{(X_n,Y_n,Z_n)\}$
with transition matrix $Q_\theta$.
Then the likelihood ratio test statistic is the difference:
\begin{equation}
    \label{eq_18}
    \Delta_n = 2 \left\{     \max_{\theta \in \Theta} L_n(X_{-k+1}^n,Y_{-k+1}^n,Z_{-k+1}^n;\theta) - \max_{\phi \in \Phi} L_n(X_{-k+1}^n,Y_{-k+1}^n,Z_{-k+1}^n;h(\phi))   \right\}.
\end{equation}

For our purposes, a key observation is that the statistics $\Delta_n$
is exactly equal to $2n$ times the plug-in estimator 
$\hat{I}_n^{(k)} (\Xp \to \Yp \| \Zp).$
Proposition~\ref{prop_3.5} is proved in Appendix~\ref{appx_proof_prop_3.5}.

\begin{proposition}
    \label{prop_3.5}
    Under the assumptions of Theorem~\ref{thmX},
	 and in the notation of this section:
    \begin{equation*}
        \Delta_n = 2 n \hat{I}_n^{(k)} (\Xp \to \Yp \| \Zp).
    \end{equation*}
\end{proposition}

Under the null hypothesis, that is, when 
$I(\Bar{Y}_0;\Bar{X}_{-k}^0|\Bar{Y}_{-k}^{-1},\Bar{Z}_{-k}^0) =0$, 
Theorem~\ref{thmX} states that the distribution of $\Delta_n$ 
is approximately $\chi^2$ with $\ell^k t^{k+1} (m^{k+1}-1) (\ell-1)$ 
degrees of freedom. Importantly,
this limiting distribution does {\em not} depend on the distribution 
of the underlying process $\{(X_n,Y_n,Z_n)\}$, 
except through the alphabet sizes $m, \ell, t$ and the memory length $k$. 
Therefore, as with classical likelihood-ratio tests,
we can decide whether or not the data offer strong enough evidence 
to reject the null hypothesis by examining the value of $\Delta_n$. 
In particular, if the $p$-value given by the probability of 
the tail $[\Delta_n,\infty)$ of the asymptotic $\chi^2$ distribution is below a certain threshold $\alpha$, then the null hypothesis of causal conditional influence can be rejected at the significance level $\alpha$.

\newpage

\section{Results on simulated data}
\label{s:simulations}

We examine the performance of the hypothesis testing
framework described in Section~\ref{sec_hypothesis_test}
on a simulated data set. As mentioned in the Introduction, 
we also take this opportunity to illustrate the performance
of the simpler temporal causality test developed in~\cite{skoularidou-K:16}
where no potential confounder was present.
Indeed, in the absence of a confounding process $\Zp$,
the plug-in estimator $\hat{I}_n^{(k)}(\Xp\to\Yp\|\Zp)$
in~(\ref{eq_10}) reduces to the simpler estimator
$\hat{I}_n^{(k)}(\Xp\to\Yp)$
in~\cite{skoularidou-K:16}. Similarly,
taking the alphabet size $t=|C|$ of the $\Zp$
process to be equal to~1,
the $\chi^2$-convergence in Theorem~\ref{thmX}
reduces to the special case obtained
in~\cite[Theorem~3.2~$(ii)$]{skoularidou-K:16}.

We call the resulting test for the presence
of temporally causal influence of $\Xp$ on $\Yp$
the {\em unconditional causality} (UC) test. Similarly,
we call the test of Section~\ref{sec_hypothesis_test}
for temporally causal influence of $\Xp$ on $\Yp$
in the causal presence of $\Zp$ the
{\em conditional causality} (CC) test.

We will examine the performance of both tests
on data generated from the following
process $\{(X_n,Y_n,Z_n)\}$.
Let $\{ X_n \}$ be a second order, binary Markov chain, 
with transition probabilities,
\begin{align*}
\BBP(X_n=0|X_{n-1}=0,X_{n-2}=0) &= 0.3, 
	& \BBP(X_n=0|X_{n-1}=0,X_{n-2}=1) &= 0.6,\\
\BBP(X_n=0|X_{n-1}=1,X_{n-2}=0) &= 0.8,
	& \BBP(X_n=0|X_{n-1}=1,X_{n-2}=1) &= 0.1,
\end{align*}
and let $\{Z_n\}$ to be an independent and identically distributed 
(i.i.d.)\ sequence of Bern$(1/2)$ variables,
independent of $\{X_n\}$
Then we define
$Y_n = X_n + Z_{n-3} + W_n \; (\mathrm{mod}\; 2)$,
for all $n$, where
$\{ W_n \}$ are i.i.d.\ Bern$(p)$ 
with $p=0.01$,
independent of $\{X_n\}$ and $\{Z_n\}$.

\medskip

\noindent
{\bf UC test. }
It is easy to see that, by construction,
the bivariate process $\{(X_n,Y_n)\}$
is a second order Markov chain and,
moreover, that $\Xp$ does {\em not} 
have a temporally causal influence on $\Yp$
so the null conditional independence
hypothesis should not be rejected in this
case.

With a sample of length $N=2 \times 10^5$ from $\{(X_n,Y_n)\}$,
the results of the UC test for whether
$\Xp$ has a temporally causal influence on $\Yp$
within the class of Markov chains of order 
$k=0,1,2,3,4,$ and~$5$, are shown 
in Table~\ref{tab_hyp_I}.
We observe 
that, as expected (and also as desired),
the null is not rejected for any~$k$.
The same test was repeated with shorter
sample sizes, giving similar results.

\begin{figure}[ht!]
\centering
\includegraphics[width=0.5\textwidth]{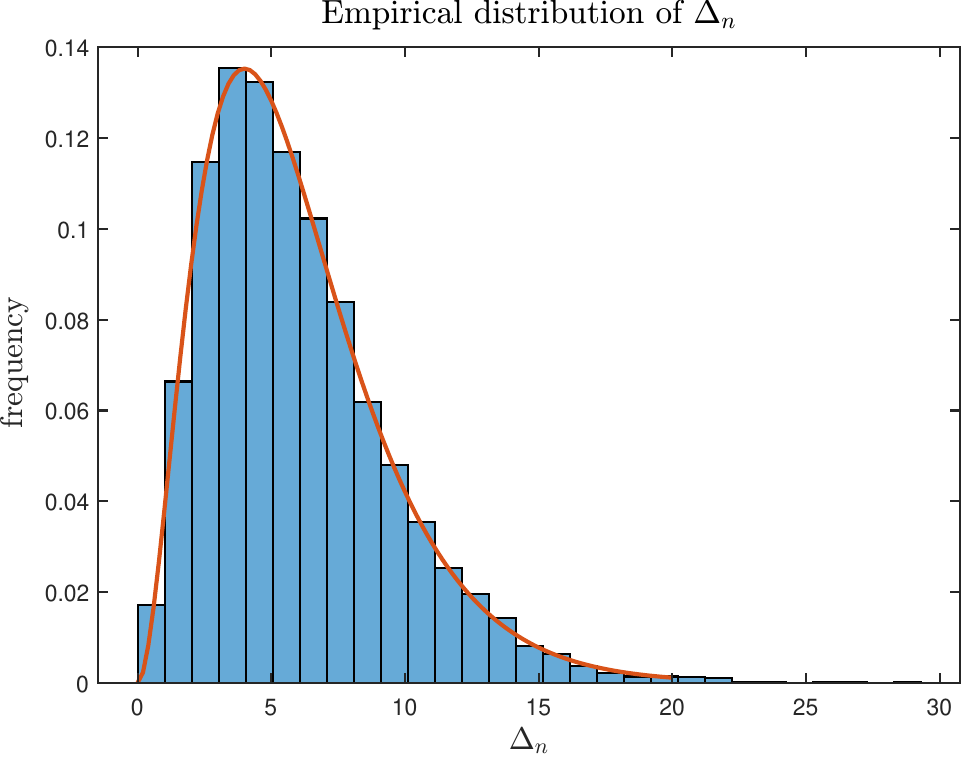}
\caption{Histogram approximation of the distribution of
the statistics
$\Delta_n$ based on 10000 repetitions of the UC test 
on samples of length $N=3\times 10^4$.
The red curve shows the
density of the theoretically predicted limiting
$\chi^2(6)$ distribution.}
\label{fig_emp_hyp_I}
\end{figure}

Using $10000$ repetitions of the test
on samples of length $N=3\times 10^4$,
the empirical distribution of the statistic $\Delta_n$
was estimated as shown in Figure~\ref{fig_emp_hyp_I} 
to be in close agreement with the theoretically 
predicted $\chi^2(6)$ limiting distribution.

\begin{table}[ht!]
\centering
\begin{tabular}{|l|l|l|l|l|l|l|}
\hline
Markov depth $k$ & 0    & 1    & 2    & 3     & 4     & 5      \\ \hline
Number of DoF    & 1    & 6    & 28   & 120   & 496   & 2016   \\ \hline
Statistic $\Delta_n$       & 0.5  & 2.2  & 28.6 & 118.1 & 491.5 & 2047.3 \\ \hline
$p$-value        & 0.46 & 0.90 & 0.43 & 0.53  & 0.55  & 0.31   \\ \hline
Reject $H_0$ at $5\%$ level & No   & No   & No   & No    & No    & No    \\ \hline
\end{tabular}
\caption{Results of the UC test
with different Markov orders $k$,
on $N=2\times 10^5$ samples.}
\label{tab_hyp_I} 
\end{table}

\medskip

\noindent
{\bf CC test. }
The joint process $\{(X_n,Y_n,Z_n)\}$ 
described above is easily seen
to be a Markov chain of order~3.
Interestingly, within the class of Markov chains of
order $k$, 
$\Xp$ {\em does} have a temporally causal
influence on $\Yp$ causally conditioned on $\Zp$ for all $k\geq 3$,
and it does {\em not} for $k\leq 2$.
Therefore, we expect the CC test to 
reject the null for $k\geq 3$, and to
not reject it for $k\leq 2$.

Using a sample length $N=2\times 10^5$,
the results of the CC test
for Markov orders $0\leq k\leq 5$ are shown in
Table~\ref{tab_hyp_II}.
As can be seen there, the null hypothesis 
is not rejected for values of $k\leq 2$, 
while it is rejected (and quite decisively)
for $k\geq 3$. As noted above, this is in perfect
agreement with the nature of the data.

\begin{table}[ht!]
\centering
\begin{tabular}{|l|l|l|l|l|l|l|}
\hline
Markov depth $k$ & 0    & 1    & 2     & 3          & 4          & 5          \\ \hline
Number of DoF          & 2    & 24   & 224   & 1920       & 15872      & 129024     \\ \hline
Statistic $\Delta_n$   & 0.6  & 13.3 & 228.2 & $2.5\times 10^5$ & $2.5\times 10^5$ 
	& $1.9\times 10^5$ \\ \hline
$p$-value      & 0.74 & 0.96 & 0.41  & $<10^{-4}$ & $<10^{-4}$ & $<10^{-4}$ \\ \hline
Reject $H_0$ at 5\% level & No   & No   & No    & Yes        & Yes        & Yes        \\ \hline
\end{tabular}
\caption{Results of the CC test 
with different Markov orders $k$,
on $N = 2\times 10^5$ samples.}
\label{tab_hyp_II}
\end{table}

\begin{figure}[ht!]
\centering
\includegraphics[width=0.83\textwidth]{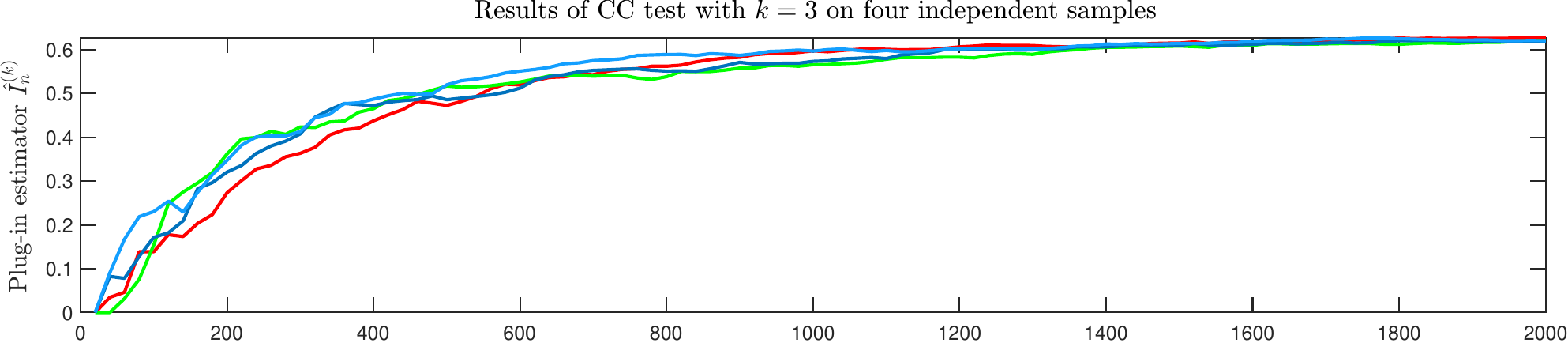}
\includegraphics[width=0.835\textwidth]{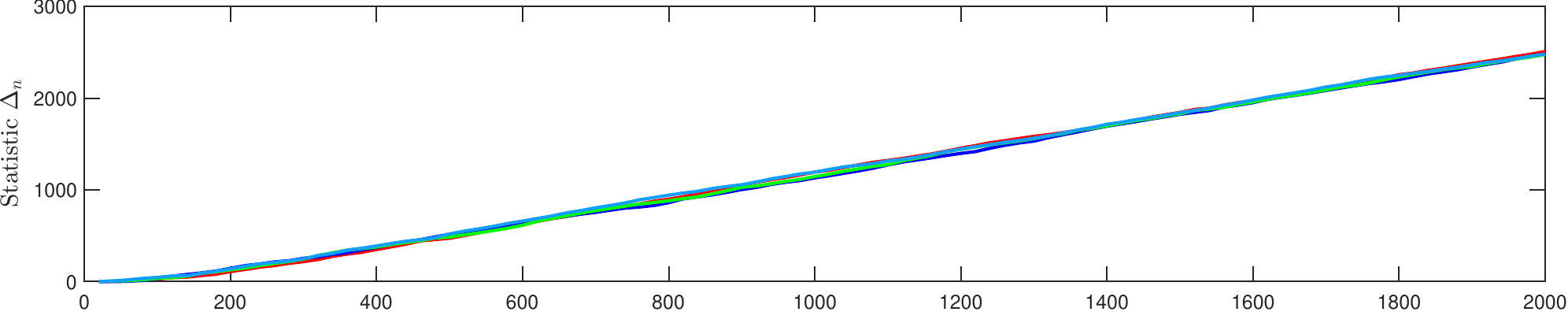}
\includegraphics[width=0.82\textwidth]{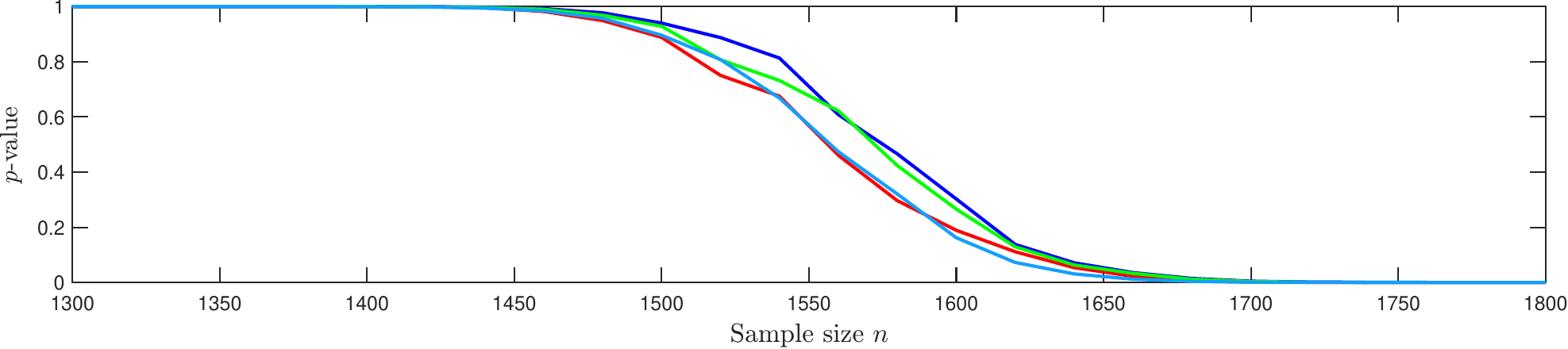}
\caption{Results of four independent repetitions of the
CC test with Markov order $k=3$. Plotted are the values
of the plug-in estimator $\hat{I}_n^{(k)}(\Xp\to\Yp\|\Z)$,
the statistic $\Delta_n$, and the corresponding $p$-value,
as functions of the sample size $n$.
The $p$-value quickly decreases towards zero,
correctly indicating the presence of temporally 
causal influence of $\Xp$ on $\Yp$ causally conditioned on $\Zp$.}
\label{fig:hyp_II_D=3}
\end{figure}

\begin{figure}[ht!]
\centering
\includegraphics[width=0.8\textwidth]{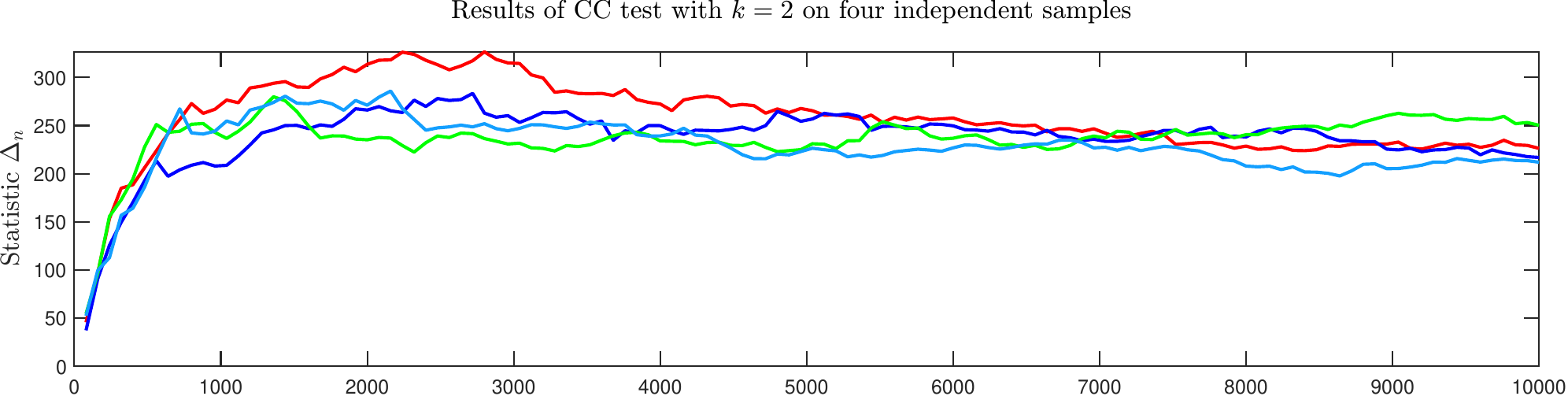}
\includegraphics[width=0.8\textwidth]{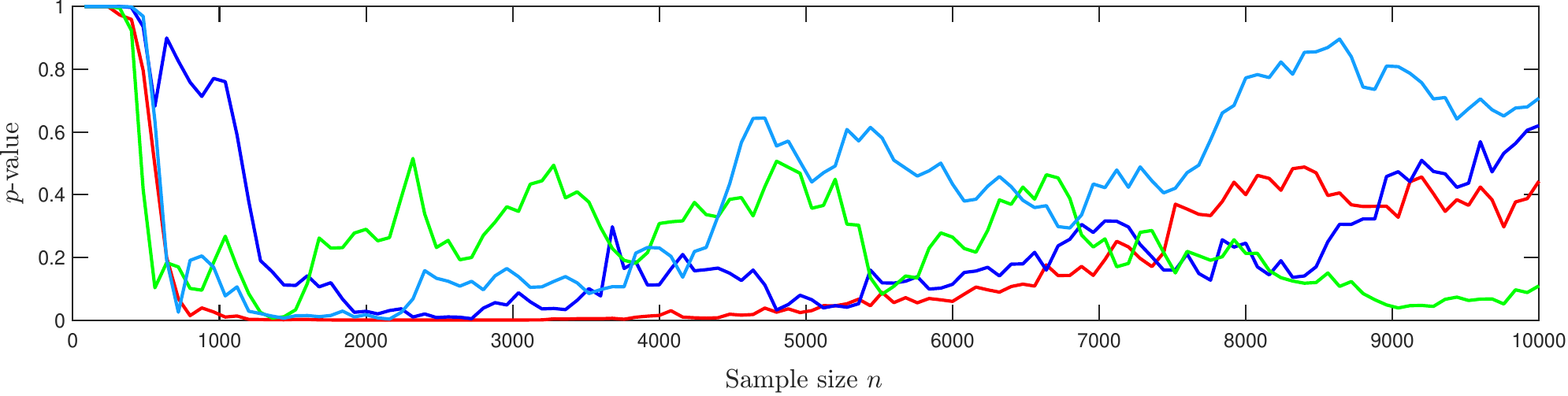}
\caption{Results of four independent repetitions of the
CC test with Markov order $k=2$.~Plotted are the values
of the statistic $\Delta_n$ and the corresponding $p$-value
as functions of the sample size~$n$.
The $p$-value stays mostly away from zero,
and the conditional independence hypothesis is not rejected
as indeed there is no temporally causal influence of $\Xp$
on $\Yp$ causally conditioned on $\Zp$ present in the data.}
\label{fig:hyp_II_D=2}
\end{figure}

The drastically different behavior
of the CC test between the two `boundary' cases
is illustrated in Figures~\ref{fig:hyp_II_D=3} and~\ref{fig:hyp_II_D=2}.
Figure~\ref{fig:hyp_II_D=3} shows the behavior of 
the plug-in 
estimator $\hat{I}_n^{(k)}(\Xp\to\Yp\|\Zp)$,
the statistic $\Delta_n$,
and the corresponding $p$-value,
as the sample
size increases from $n=1$ to $n=10000$, in four independent
repetitions of the same experiment with $k=3$. 
As predicted by Theorem~\ref{thmN}, the plug-in 
estimator $\hat{I}_n^{(k)}(\Xp\to\Yp\|\Zp)$
converges to the true value
$I(\Xp\to\Yp\|\Zp)\approx 0.6103$
as $n$ increases, and the value of the 
test statistic grows with $n$.
As a result, for $n$ between roughly $1500$
and $1650$
the corresponding $p$-value drops
sharply to values very close 
to zero, indicating that the conditional
independence hypothesis be (correctly) rejected,
as indeed in this case $\Xp$ does have a temporally causal
influence on $\Yp$ causally conditioned on $\Zp$.

By contrast, when the Markov order $k=2$, 
the results shown Figure~\ref{fig:hyp_II_D=3} 
indicate very different behavior.
The statistic $\Delta_n$ neither increases
nor does it converge to a specific value with~$n$,
in agreement with Theorem~\ref{thmX} which predicts
that it converges to a non-degenerate $\chi^2$
distribution. Moreover,
the $p$-value
stays significantly away from zero in almost all cases,
indicating that the conditional independence
hypothesis cannot be rejected in this case,
as indeed this conditional independence is present
in the data.

\newpage

$\;$

\newpage

\section{Experimental results on neural spike trains}
\label{s:spikes}

Here we revisit a data set consisting of neural 
spike trains recorded from awake behaving monkeys.
The objective of the original study was to investigate 
the neural mechanisms that underlie spatial attention;
see~\cite{gregoriou:09,gregoriou:12} for detailed descriptions
of the relevant experiments and the data acquisition process.
Specifically, we examine simultaneous recordings of spike trains
of individual neurons in the frontal eye field (FEF),
an area in the prefrontal cortex,
and area V4 in the visual cortex. The spike trains are 
quantized into 1~millisecond (ms) bins, 
to form discrete-time, binary-valued time series.

The data set consists 
of single neurons' responses, recorded using multiple 
electrodes across 138 experimental sessions. For the purpose of 
this analysis, we only consider pairs of neurons across FEF and 
V4 that were recorded concurrently during a single experimental session.
In each session, 8 different task epochs were repeated in a 
specific order (onset of fixation, onset of visual stimuli, 
onset of instruction cue on which stimulus to attend, 
onset of distractor color change, onset of target color change, 
etc.) and each sequence was considered as an experimental trial. 
Neuronal responses from different trials within a session were 
concatenated and the resulting spike trains had variable lengths in
the order of $5\times 10^6$~ms.
The different task epochs were also
recorded simultaneously with the spike
trains, enabling the generation of a binary
time series of `events' at a resolution of 1~ms, 
with a `1' 
signifying the `cue onset' epoch, and
`0' signifying all other epochs; 
see~\cite{gregoriou:09,gregoriou:12} 
for an extensive discussion of the
experimental procedure.

Previous studies suggested that discrete-time estimates of directional 
coupling using temporal binning of spike data can result in estimation 
biases~\cite{mijatovic:21}. This was mainly evident using time bin widths 
of 0.1~s or larger and spike train lengths in the order of $10^3$~s 
or shorter. We chose a fine discretization of 1~ms, that is two orders 
of magnitude smaller, which takes into account the minimum refractory 
period of neuronal firing and results in no loss of information. Given 
that conduction and synaptic delays for monosynaptic connections 
in the brain are in the 
order of 5-10~ms, the enhanced temporal resolution that preserves the 
statistics of the spike train is of utmost importance to detect 
significant dependencies in real data from spiking activity. Moreover, 
the use of such short time bins together with long spike trains in 
the order of $5\times 10^6$~ms in our study, should result in minimal 
estimation biases. It has also been noted that since structure 
in spike trains is often in the order of tens or several hundred 
milliseconds, using such a fine binning will result in impractically 
large history dependencies~\cite{spinney:17}. However, in our study, 
we are looking at dependencies between monosynaptically connected 
brain regions with latencies of information transfer in the order 
of 10~ms~\cite{gregoriou:09}. Larger temporal history dependencies in 
spike trains across brain regions would be hard to interpret biologically 
and are beyond the scope of this work.

Broadly speaking, our goal here is to examine
whether neurons in one of the two brain regions
(V4 and FEF) have a temporally causal influence
on neurons in the other. To that end,
we employ the hypothesis tests described in the 
previous sections in three different settings.

\medskip

\noindent
{\bf A. Causal influence V4$\to$FEF or FEF$\to$V4. }
First we examine the presence of temporally causal
influences between individual neurons,
based on the unconditional causality test UC outlined in
Section~\ref{s:simulations}, with $k=2$. Specifically,
we consider 344 pairs of simultaneously recorded spike trains
$\{(X_n,Y_n)\}$, with $\{X_n\}$ being the
spike train of a neuron in one region
and $\{Y_n\}$ the spike train
of a simultaneously recorded neuron in the other region. 

From these 344 hypothesis tests, 
85 pairs of neurons were identified
where temporally causal influence
was present, with the null being rejected
at a 5\% significance level.
In most cases, whenever causal influence
$\Xp\to\Yp$ was detected, the same
was found in the reverse direction, $\Yp\to\Xp$,
indicating that, typically, causal 
influence occurs in both directions 
simultaneously between FEF and V4 .

\begin{figure}[ht!]
\centering
\includegraphics[width=0.82\textwidth]{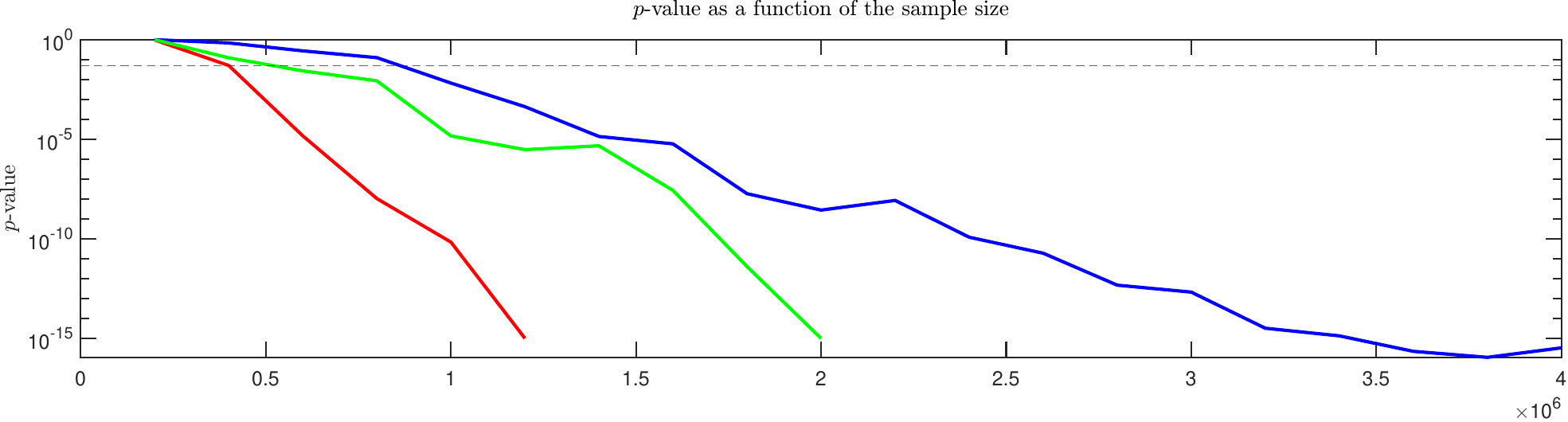}
\includegraphics[width=0.8\textwidth]{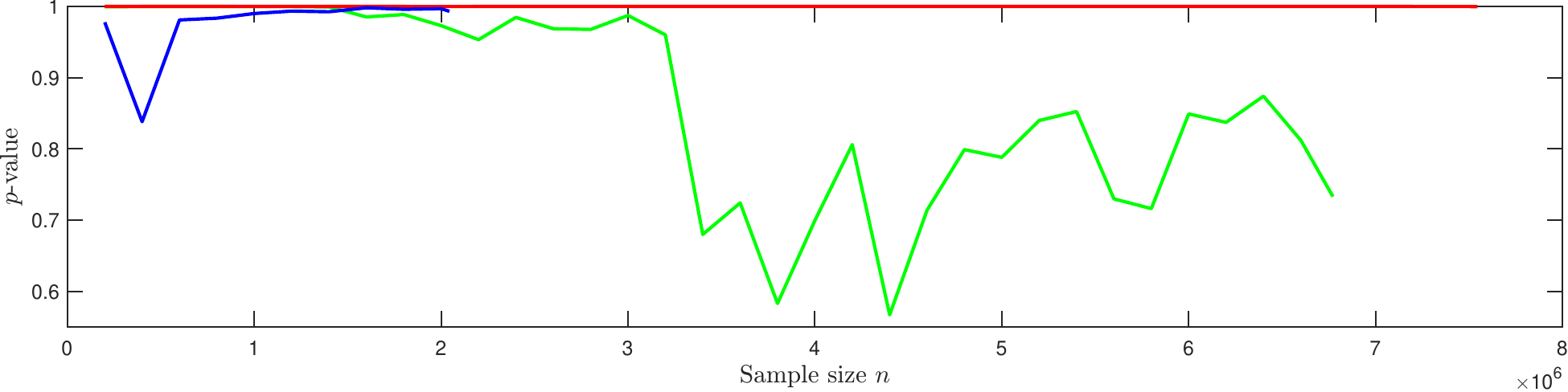}
\caption{The $p$-value of the UC test between pairs of
neurons, as a function of the sample size.
The top plot shows the results in three cases
of pairs of neurons where the null was rejected,
and the bottom plot shows three cases where it was not.
The dotted horizontal line shows the $5\%$ significance
level.}
\label{fig:exp_A}
\end{figure}

The top plot in Figure~\ref{fig:exp_A} shows
the behavior of the $p$-value as a function
of the sample size, for three pairs of neurons
where causal influence was detected,
and the bottom plot shows three corresponding
examples where it was not. Note that the
individual
graphs have different lengths as the spike
trains considered have varying durations.
In almost all of the 344 tests performed we observed
a clear dichotomy between two distinct 
behaviors.
The $p$-value would either decrease quickly
to values very close to zero, or it would
stay very close to~1.

\medskip

\noindent
{\bf B. Causal influence conditional on stimuli. }
Next we consider the same 344 cases as in~A., except
now we allow for causal conditioning on the
`events' time series $\{Z_n\}$ described above.
The conditional causality test CC 
of
Sections~\ref{sec_hypothesis_test}
and~\ref{s:simulations}
was used with $k=2$.  
This time, temporally causal influence was identified
in only 20 cases at a $5\%$ significance level.
This clearly suggests a very significant
confounding effect of the events 
to both the FEF and V4 regions.
Moreover, since 20/344 is approximately
$5.8\%$, even the cases where causal influence appears
to be present, the results are not necessarily reliable.
Also, as in~A., the $p$-values obtained were either very
small, typically $<10^{-3}$, or almost identically equal
to~1.

%

\medskip

\noindent
{\bf C. Causal influence conditional on another spike train. }
Finally, we consider a scenario similar to that in~B., 
except now the potentially confounding process $\{Z_n\}$ 
is taken to be another spike train from a V4 or FEF neuron.
Again, the CC test was performed with $k=2$.
In 228 hypothesis tests, only 4 cases 
were found where temporally causal influence was present,
that is, with the null rejected at a 5\% significance level. 
Once again, as $4/228\approx 1.6\%$, the cases
where causal influence was identified do not represent
reliable findings.

Here, the presence of a third neuron's spike train appears 
to have an essentially
overwhelming confounding effect.
This finding is not surprising given the much denser intrinsic synaptic 
connections among neurons within each brain region compared to between 
region synaptic density~\cite{anderson:11}. Based on this, it is expected 
that within each region the majority of neuronal responses are mainly 
affected by other neurons within the same region. Substantially fewer 
connections are found between FEF and V4~\cite{anderson:11}. However, 
the absence of a causal influence between pairs of neurons when other 
spike trains are taken into account may also reflect the fact 
that stimuli impact many neurons within both 
FEF and V4 simultaneously.

\medskip

\noindent
{\bf Concluding remarks. }
The results in this section are mostly illustrative 
of the hypothesis tests introduced
in earlier sections.
Further analysis would be required 
in order to provide conclusive
biologically relevant findings, specifically,
to fully characterize and identify the causal influences 
in the FEF and V4 regions.
However, our analysis does indicate that the hypothesis 
testing framework described in this paper 
can be a useful exploratory tool in this direction.
Moreover, the UC and CC tests can be applied in very 
broad range of applications beyond neuroscience,
including time series from meteorology and finance;
see, e.g., some of the relevant data sets examined
in~\cite{ctw-isit:21,lungu-arxiv:22,lungu-pap-K:22,BCT-JRSSB:22}.

Returning to spike train data, we recall that 
there is a latent time of about 10~ms for a signal 
to travel from FEF 
to V4 or vice versa~\cite{gregoriou:09}.
Therefore, it is natural to ask whether 
the same instances of temporally causal influence
identified here are still present when the $\Yp$ process
is shifted by 10~ms.
Indeed, in 74 out of these 85 pairs of neurons,
the presence of causal influence observed persists after 
this shift.

\newpage

\section*{Acknowledgements}
The experimental spike data analyzed in this study were 
obtained in the Lab of Robert Desimone (NIMH, NIH, U.S.A.). 
The help of Stephen J.\ Gotts in the acquisition of these 
data is gratefully acknowledged. We are also grateful to the
Associate Editor and the
two anonymous reviewers for their useful comments and suggestions.

\appendix 
\section{Appendix}
\subsection{Proof of Proposition \ref{prop_irr,aper}}
\label{app:proposition}

We find it convenient to first establish
the following ``stationary'' version of 
Proposition~\ref{prop_irr,aper}.

\begin{proposition}
\label{prop_stationary}
If $\{(X_n, Y_n, Z_n)\}$ is a stationary Markov chain of order no 
larger than $k \geq 1$, then:
    \begin{enumerate}
        \item[$(i)$] The causal conditional entropy 
	rate $H(\Yp \| \Zp )$ exists.
        \item[$(ii)$] The CCDI rate $I(\Xp \to \Yp \| \Zp )$ exists 
	and it equals,
        \begin{equation*}
            I(\Xp \to \Yp \| \Zp )
	= H(\Yp \| \Zp ) -H(Y_0 | Y_{-k}^{-1},X_{-k}^0, Z_{-k}^0).
        \end{equation*}
        \item[$(iii)$] If $ \{(Y_n, Z_n) \}$ is also a Markov chain of order 
	no larger than $k$, then:
        \begin{equation*}
            I(\Xp \to \Yp \| \Zp ) = I(Y_0 ; X_{-k}^0 | Y_{-k}^{-1}, Z_{-k}^0).
        \end{equation*}
    \end{enumerate}
\end{proposition}

\noindent 
{\sc Proof. }
Notice that,
\begin{equation}
         \frac{1}{n} H(Y_1^n \| Z_1^n) 
	=  \frac{1}{n} \sum_{i=1}^n  H(Y_i | Y_1^{i-1}, Z_1^i),
	\label{eq:cHrate}
\end{equation}
where, by stationarity, the summands satisfy,
$$H(Y_i | Y_1^{i-1}, Z_1^i)=
H(Y_0 | Y_{-i+1}^{-1}, Z_{-i+1}^0),$$
so they are nonincreasing in $i$ 
and nonnegative. Therefore, they necessarily
converge to a nonnegative limit, and the
Ces\`{a}ro averages in~(\ref{eq:cHrate})
converge to the same limit. This proves~$(i)$.

For part $(ii)$, first we write,
\begin{align}
        H(Y_1^n \| X_1^n, Z_1^n)  
	&= \sum_{i=1}^n H(Y_i | Y_1^{i-1},X_1^i, Z_1^i) 
	\nonumber \\ 
	&= \sum_{i=1}^k H(Y_i | Y_1^{i-1},X_1^i, Z_1^i) 	
	+ \sum_{i=k+1}^n H(Y_i | Y_1^{i-1},X_1^i, Z_1^i),
	\label{eq_prop_ii_a} 
\end{align}
so that, by the Markov property and stationarity,
\begin{align}
        H(Y_1^n \| X_1^n, Z_1^n)  
        &= \sum_{i=1}^k H(Y_i | Y_1^{i-1},X_1^i, Z_1^i) 
	+ \sum_{i=k+1}^n H(Y_i | Y_{i-k}^{i-1},X_{i-k}^i, Z_{i-k}^i) 
	\label{eq_prop_ii_ab}\\ 
	&= \sum_{i=1}^k H(Y_i | Y_1^{i-1},X_1^i, Z_1^i) 
	+ (n-k) H(Y_0 | Y_{-k}^{-1},X_{-k}^0, Z_{-k}^0).
	\label{eq_prop_ii_b} 
\end{align}
Dividing by $n$ and taking the limit $n\to\infty$, the first
term vanishes since the first sum in~(\ref{eq_prop_ii_b})
is bounded, and we obtain,
\bqq
\lim_{n\to\infty}\frac{1}{n}
        H(Y_1^n \| X_1^n, Z_1^n)  
	=H(Y_0 | Y_{-k}^{-1},X_{-k}^0, Z_{-k}^0).
\label{eq:c2Hrate}
\eqq
Recalling the definitions of the CCDI~(\ref{eq:CCDI})
and CCDI rate~(\ref{eq_3}),
and combining them with~(\ref{eq:c2Hrate}) and the
result of part~$(i)$, yields~$(ii)$.

Finally, if $\{(Y_n, Z_n) \}$ is also a Markov chain of order 
no larger than $k$, then proceeding as in~(\ref{eq_prop_ii_a})
and~(\ref{eq_prop_ii_b}),
we have,
$$
H(Y_1^n \| Z_1^n) 
        = \sum_{i=1}^k H(Y_i | Y_1^{i-1}, Z_1^i) 
	+ (n-k) H(Y_0 | Y_{-k}^{-1}, Z_{-k}^0),
$$ 
so that $H(\Yp\|\Zp)=H(Y_0 | Y_{-k}^{-1}, Z_{-k}^0).$
Substituting this into the result of part~$(ii)$,
yields,
$$
I(\Xp \to \Yp \| \Zp )  
= H(Y_0 | Y_{-k}^{-1}, Z_{-k}^0) -H(Y_0 | Y_{-k}^{-1},X_{-k}^0, Z_{-k}^0)
= I(Y_0 ; X_{-k}^0 | Y_{-k}^{-1}, Z_{-k}^0),
$$
completing the proof.
\qed

\noindent
{\sc Proof of Proposition \ref{prop_irr,aper}. }
The ergodicity of the joint process
$\{(X_n, Y_n, Z_n)\}$ implies that it is
asymptotically mean stationary (AMS)~\cite{GraKie:Asymptotically:1980},
and hence so is the bivariate process
$\{(Y_n, Z_n)\}$, as a 
stationary coding of $\{(X_n, Y_n, Z_n)\}$~\cite{gray-90:book}.
Arguing as in the proof of 
Proposition~\ref{prop_stationary}~$(i)$,
the existence of $H(\Yp \| \Zp )$ 
and the expression in~(\ref{eq:ccH})
follow from
the general results in~\cite{gray-90:book}.
This proves~$(i)$.

Again arguing as in the proof of 
Proposition~\ref{prop_stationary}~$(ii)$,
from~(\ref{eq_prop_ii_ab})
we have,
\begin{align}
H(Y_1^n \| X_1^n, Z_1^n)
&= 
	\sum_{i=1}^k [ H(Y_i | Y_1^{i-1},X_1^i, Z_1^i) 
	- H(Y_i | Y_{i-k}^{i-1},X_{i-k}^i, Z_{i-k}^i) ]
	\nonumber\\
&
	\quad
	+ \sum_{i=1}^n H(Y_i | Y_{i-k}^{i-1},X_{i-k}^i, Z_{i-k}^i).
        \label{prop2_ii_entropy_proof}
\end{align}
The first sum above is bounded over $n$, 
and by the
ergodicity of $\{(X_n,Y_n,Z_n)\}$
and the fact that conditional entropy is continuous over
finite alphabets,
the summands
in the second sum satisfy,
$H(Y_i | Y_{i-k}^{i-1},X_{i-k}^i, Z_{i-k}^i) 
\to H(\Bar{Y}_0 | \Bar{Y}_{-k}^{-1},\Bar{X}_{-k}^0,\Bar{Z}_{-k}^0)$ 
as $i \to \infty$.
Dividing~\eqref{prop2_ii_entropy_proof} by $n$ and letting $n \to \infty$, 
we therefore have,
\begin{equation*}
    \frac{1}{n} \sum_{i=1}^n H(Y_i | Y_{i-k}^{i-1},X_{i-k}^i, Z_{i-k}^i)  \to H(\Bar{Y}_0 | \Bar{Y}_{-k}^{-1},\Bar{X}_{-k}^0, \Bar{Z}_{-k}^0).
\end{equation*}
Combining this with the result of part~$(i)$ and
the definition of the CCDI rate proves part~$(ii)$.

Now suppose $ \{(Y_n, Z_n) \}$ is also a Markov chain of 
order $k$. Since $\{(X_n, Y_n, Z_n)\}$ is 
irreducible and aperiodic, hence ergodic,
and since $(Y_n, Z_n)$ is trivially a function of $(X_n, Y_n, Z_n)$, 
the chain $ \{(Y_n, Z_n) \}$ is also irreducible and aperiodic.
Arguing as in~$(ii)$, we first have,
$$H(Y_1^n \| Z_1^n) 
= \sum_{i=1}^k [ H(Y_i | Y_1^{i-1}, Z_1^i) - H(Y_i | Y_{i-k}^{i-1}, Z_{i-k}^i) ]+ \sum_{i=1}^n H(Y_i | Y_{i-k}^{i-1}, Z_{i-k}^i),
$$
from which it follows that, as $n\to\infty$,
\begin{equation*}
    \frac{1}{n} H(Y_1^n \| Z_1^n) \to H(\Bar{Y}_0 | \Bar{Y}_{-k}^{-1}, \Bar{Z}_{-k}^0).
\end{equation*}
Therefore, in this case $I(\Xp \to \Yp \| \Zp )$ simplifies to,
\begin{equation*}
    \begin{split}
        I(\Xp \to \Yp \| \Zp )  &= H(\Yp \| \Zp ) -H(\Bar{Y}_0 | \Bar{Y}_{-k}^{-1},\Bar{X}_{-k}^0, \Bar{Z}_{-k}^0) \\
        &= H(\Bar{Y}_0 | \Bar{Y}_{-k}^{-1}, \Bar{Z}_{-k}^0) - H(\Bar{Y}_0 | \Bar{Y}_{-k}^{-1},\Bar{X}_{-k}^0, \Bar{Z}_{-k}^0) \\
        &= I(\Bar{Y}_0 ; \Bar{X}_{-k}^0 | \Bar{Y}_{-k}^{-1}, \Bar{Z}_{-k}^0),
    \end{split}
\end{equation*}
establishing~$(iii)$ and completing the proof. \qed

\subsection{Proof of Theorem~\ref{thmN}}
\label{proofthmN}

We begin by expressing the plug-in 
estimator $\hat{I}_n^{(k)} (\Xp \to \Yp\| \Zp)$ 
in a more complex-looking but more manageable form.
Recall the definitions of the empirical 
distribution $\hat{P}_{X_{-k}^0,Y_{-k}^0,Z_{-k}^0,n}$ and of 
the plug-in estimator $\hat{I}_n^{(k)} (\Xp \to \Yp\| \Zp)$, in 
equations~\eqref{eq_9} and~\eqref{eq_10} respectively.
Expanding the definitions we have,
\begin{align}
\hat{I}_n^{(k)} (\Xp \to \Yp\| \Zp) 
&= 
	I(\hat{Y}_0 ; \hat{X}_{-k}^0|\hat{Y}_{-k}^{-1},\hat{Z}_{-k}^0 )
	\nonumber\\
&=    
	H(\hat{Y}_0 | \hat{Y}_{-k}^{-1}, \hat{Z}_{-k}^0) 
	- H(\hat{Y}_0 | \hat{X}_{-k}^0, \hat{Y}_{-k}^{-1}, \hat{Z}_{-k}^0)
	\nonumber\\ 
&= 
	- \sum_{b_0^k, c_0^k } \hat{P}_{Y_{-k}^0,Z_{-k}^0,n} (b_0^k, c_0^k) 
	\log \left( \frac{\hat{P}_{Y_{-k}^0,Z_{-k}^0,n} (b_0^k, c_0^k)}
	{\hat{P}_{Y_{-k}^{-1},Z_{-k}^0,n} (b_0^{k-1}, c_0^k)} \right) 
	\nonumber\\
& 
	\quad\, + \sum_{a_0^k, b_0^k, c_0^k }  
	\hat{P}_{X_{-k}^0,Y_{-k}^0,Z_{-k}^0,n} (a_0^k, b_0^k, c_0^k) 
	\log \left( \frac{\hat{P}_{X_{-k}^0,Y_{-k}^0,Z_{-k}^0,n} 
	(a_0^k, b_0^k, c_0^k)}{\hat{P}_{X_{-k}^0,Y_{-k}^{-1},Z_{-k}^0,n} 
	(a_0^k, b_0^{k-1}, c_0^k)} \right),
	\nonumber
\end{align}
and, omitting the arguments $(a_0^k, b_0^k, c_0^k)$
of the probability mass functions for the sake of visual
clarity, we can express,
\begin{align}
\hat{I}_n^{(k)} 
& 
	(\Xp \to \Yp\| \Zp) 
	\nonumber\\
&= 
	\sum_{a_0^k, b_0^k, c_0^k }  
	\hat{P}_{X_{-k}^0,Y_{-k}^0,Z_{-k}^0,n} \log 
	\left( \frac{\hat{P}_{Y_{-k}^{-1},Z_{-k}^0,n}  
	\hat{P}_{X_{-k}^0,Y_{-k}^0,Z_{-k}^0,n} }{\hat{P}_{Y_{-k}^0,Z_{-k}^0,n}
	\hat{P}_{X_{-k}^0,Y_{-k}^{-1},Z_{-k}^0,n} } \right)
	\nonumber\\
&= 
	\sum_{a_0^k, b_0^k, c_0^k} \hat{P}_{X_{-k}^0,Y_{-k}^0,Z_{-k}^0,n} 
	\log \left( \frac{P_{\Bar{Y}_{-k}^{-1},\Bar{Z}_{-k}^0} 
	P_{\Bar{X}_{-k}^0,\Bar{Y}_{-k}^0,\Bar{Z}_{-k}^0} }
	{P_{\Bar{Y}_{-k}^0,\Bar{Z}_{-k}^0} 
	P_{\Bar{X}_{-k}^0,\Bar{Y}_{-k}^{-1},\Bar{Z}_{-k}^0} } \right)                   \nonumber\\
& 
	\qquad + \sum_{a_0^k, b_0^k, c_0^k } 
	\Bigg\{ \hat{P}_{X_{-k}^0,Y_{-k}^0,Z_{-k}^0,n} \log \Bigg( 
	\frac{\hat{P}_{Y_{-k}^{-1},Z_{-k}^0,n} 
	\hat{P}_{X_{-k}^0,Y_{-k}^0,Z_{-k}^0,n} }{\hat{P}_{Y_{-k}^0,Z_{-k}^0,n}
	\hat{P}_{X_{-k}^0,Y_{-k}^{-1},Z_{-k}^0,n} } \times
	\frac{P_{\Bar{Y}_{-k}^0,\Bar{Z}_{-k}^0} 
	P_{\Bar{X}_{-k}^0,\Bar{Y}_{-k}^{-1},\Bar{Z}_{-k}^0} }
	{P_{\Bar{Y}_{-k}^{-1},\Bar{Z}_{-k}^0} 
	P_{\Bar{X}_{-k}^0,\Bar{Y}_{-k}^0,\Bar{Z}_{-k}^0} } \Bigg)  \Bigg\}.
        \nonumber
\end{align}
Substituting the definition
of the empirical probability mass functions
and rearranging,
yields,
\begin{align*}
\hat{I}_n^{(k)} 
& 
	(\Xp \to \Yp\| \Zp) 
	\nonumber\\
&= 
	\sum_{a_0^k, b_0^k, c_0^k } \Bigg[ \left(\frac{1}{n} 
	\sum_{i=1}^n 
	\mathbb{I}_{ \{ X_{i-k}^i=a_0^k,Y_{i-k}^i=b_0^k,Z_{i-k}^i=c_0^k \} }  
	\right)                                               
        \log \left( 
	\frac{ P_{\Bar{X}_{-k}^0,\Bar{Y}_0|\Bar{Y}_{-k}^{-1},\Bar{Z}_{-k}^0} }
	{P_{\Bar{Y}_0|\Bar{Y}_{-k}^{-1},\Bar{Z}_{-k}^0} 
	P_{\Bar{X}_{-k}^0|\Bar{Y}_{-k}^{-1},\Bar{Z}_{-k}^0} } \right)  \Bigg]
	\nonumber\\
& 
	\qquad + 
	D \big(  \hat{P}_{Y_{-k}^{-1},Z_{-k}^0,n} 
	\| P_{\Bar{Y}_{-k}^{-1},\Bar{Z}_{-k}^0} \big) 
	+ D \big( \hat{P}_{X_{-k}^0,Y_{-k}^0,Z_{-k}^0,n}\| 
	P_{\Bar{X}_{-k}^0,\Bar{Y}_{-k}^0,\Bar{Z}_{-k}^0}    \big) 
	\nonumber\\ 
& 
	\qquad - 
	D \big( \hat{P}_{Y_{-k}^0,Z_{-k}^0,n} \| 
	P_{\Bar{Y}_{-k}^0,\Bar{Z}_{-k}^0}  \big)   
	- D \big( \hat{P}_{X_{-k}^0,Y_{-k}^{-1},Z_{-k}^0,n} 
	\| P_{\Bar{X}_{-k}^0,\Bar{Y}_{-k}^{-1},\Bar{Z}_{-k}^0}   \big),
\end{align*}
so that,
\begin{align}
\hat{I}_n^{(k)} 
& 
	(\Xp \to \Yp\| \Zp) 
	\nonumber\\
&=  
	\frac{1}{n} \sum_{i=1}^n    \log \left( 
	\frac{ P_{\Bar{X}_{-k}^0,\Bar{Y}_0|\Bar{Y}_{-k}^{-1},\Bar{Z}_{-k}^0} 
	(X_{i-k}^i, Y_i | Y_{i-k}^{i-1}, Z_{i-k}^i)}
	{P_{\Bar{Y}_0|\Bar{Y}_{-k}^{-1},\Bar{Z}_{-k}^0} 
	(Y_i|Y_{i-k}^{i-1}, Z_{i-k}^i) 
	P_{\Bar{X}_{-k}^0|\Bar{Y}_{-k}^{-1},\Bar{Z}_{-k}^0} 
	(X_{i-k}^i |Y_{i-k}^{i-1}, Z_{i-k}^i)} \right)
	\nonumber\\
& 
	\qquad + 
	D \big(  \hat{P}_{Y_{-k}^{-1},Z_{-k}^0,n} \| 
	P_{\Bar{Y}_{-k}^{-1},\Bar{Z}_{-k}^0} \big) 
	+ D \big( \hat{P}_{X_{-k}^0,Y_{-k}^0,Z_{-k}^0,n}\| 
	P_{\Bar{X}_{-k}^0,\Bar{Y}_{-k}^0,\Bar{Z}_{-k}^0}    \big) 
	\nonumber\\
& 
	\qquad - 
	D \big( \hat{P}_{Y_{-k}^0,Z_{-k}^0,n} \| 
	P_{\Bar{Y}_{-k}^0,\Bar{Z}_{-k}^0}  \big)   
	- D \big( \hat{P}_{X_{-k}^0,Y_{-k}^{-1},Z_{-k}^0,n} \| 
	P_{\Bar{X}_{-k}^0,\Bar{Y}_{-k}^{-1},\Bar{Z}_{-k}^0}   \big). 
	\label{eq_26}
\end{align}

Next, we argue that each of the four relative entropies above, 
multiplied by $\sqrt{n}$, converges to zero a.s.\ as $n \to \infty$. 
As stated in the beginning of Section~\ref{sec_plug_in_estor_and_thms}, 
the $(k+1)$-dimensional chain $\{W_n = ( X_{n-k}^n, Y_{n-k}^n, Z_{n-k}^n)\}$ 
on $A^{k+1} \times B^{k+1} \times C^{k+1}$ is ergodic, so it satisfies 
the ergodic theorem, the central limit theorem, and the law of the iterated 
logarithm~\cite{chung:book}. 

Consider the first relative entropy in~(\ref{eq_26}), 
$D \big(  \hat{P}_{Y_{-k}^{-1},Z_{-k}^0,n} 
\| P_{\Bar{Y}_{-k}^{-1},\Bar{Z}_{-k}^0} \big)$. 
Applying a quadratic Taylor expansion to the logarithm 
in the definition of the relative entropy, gives,
\begin{align}
\sqrt{n} 
& 
	D \big(\hat{P}_{Y_{-k}^{-1},Z_{-k}^0,n}
	\|P_{\Bar{Y}_{-k}^{-1},\Bar{Z}_{-k}^0}\big)
	\nonumber\\
&= 
	-\sqrt{n} \sum_{b_0^{k-1}, c_0^k } \hat{P}_{Y_{-k}^{-1},Z_{-k}^0,n} (b_0^{k-1}, c_0^k) \log \left( \frac{P_{\Bar{Y}_{-k}^{-1},\Bar{Z}_{-k}^0} (b_0^{k-1}, c_0^k)}{\hat{P}_{Y_{-k}^{-1},Z_{-k}^0,n} (b_0^{k-1}, c_0^k)} \right)    \nonumber 
        \\&= -\sqrt{n} \sum_{b_0^{k-1}, c_0^k } \Bigg\{ \hat{P}_{Y_{-k}^{-1},Z_{-k}^0,n} (b_0^{k-1}, c_0^k) \times \nonumber
        \\ &  \hspace{0.5in} 
	\times \left[ \left( \frac{P_{\Bar{Y}_{-k}^{-1},\Bar{Z}_{-k}^0} (b_0^{k-1}, c_0^k)}{\hat{P}_{Y_{-k}^{-1},Z_{-k}^0,n} (b_0^{k-1}, c_0^k)}-1 \right) -\frac{1}{2}\left( \frac{P_{\Bar{Y}_{-k}^{-1},\Bar{Z}_{-k}^0} (b_0^{k-1}, c_0^k)}{\hat{P}_{Y_{-k}^{-1},Z_{-k}^0,n} (b_0^{k-1}, c_0^k)}-1 \right)^2\frac{1}{\xi_n (b_0^{k-1}, c_0^k)^2} \right]  \Bigg\},  \nonumber
\end{align}
for some $\xi_n (b_0^{k-1}, c_0^k)$ between 1 and 
$P_{\Bar{Y}_{-k}^{-1},\Bar{Z}_{-k}^0} (b_0^{k-1}, c_0^k) / 
\hat{P}_{Y_{-k}^{-1},Z_{-k}^0,n} (b_0^{k-1}, c_0^k)$.
Therefore, we can further express,
\begin{align}
\sqrt{n} 
& 
	D \big(\hat{P}_{Y_{-k}^{-1},Z_{-k}^0,n}
	\|P_{\Bar{Y}_{-k}^{-1},\Bar{Z}_{-k}^0}\big)
	\nonumber\\
        &= \frac{\sqrt{n}}{2} \sum_{b_0^{k-1}, c_0^k }  \hat{P}_{Y_{-k}^{-1},Z_{-k}^0,n} (b_0^{k-1}, c_0^k) \left( \frac{P_{\Bar{Y}_{-k}^{-1},\Bar{Z}_{-k}^0} (b_0^{k-1}, c_0^k)}{\hat{P}_{Y_{-k}^{-1},Z_{-k}^0,n} (b_0^{k-1}, c_0^k)}-1 \right)^2\frac{1}{\xi_n (b_0^{k-1}, c_0^k)^2}     \nonumber 
        \\&= \frac{\log\log n}{2\sqrt{n}} \sum_{b_0^{k-1}, c_0^k } \Bigg[  \frac{1}{ \hat{P}_{Y_{-k}^{-1},Z_{-k}^0,n} (b_0^{k-1}, c_0^k) \xi_n (b_0^{k-1}, c_0^k)^2} \times  \nonumber
        \\& \hspace{1.5in} \times \left( \hat{P}_{Y_{-k}^{-1},Z_{-k}^0,n} (b_0^{k-1}, c_0^k) - P_{\Bar{Y}_{-k}^{-1},\Bar{Z}_{-k}^0} (b_0^{k-1}, c_0^k) \right)^2 \left( \frac{n}{\log \log n} \right)  \Bigg].
\label{eq:for_bounded}
\end{align}
By the ergodic theorem, the sequences
$\{1/\hat{P}_{Y_{-k}^{-1},Z_{-k}^0,n} (b_0^{k-1}, c_0^k) \}$ 
and $\{ 1 / \xi_n (b_0^{k-1}, c_0^k)^2\}$ are both bounded a.s.\
in $n$, and the law of the iterated logarithm implies that 
the product of the next two terms,
\begin{align}
        \Big( \hat{P}_{Y_{-k}^{-1},Z_{-k}^0,n} (b_0^{k-1}, c_0^k) & - P_{\Bar{Y}_{-k}^{-1},\Bar{Z}_{-k}^0} (b_0^{k-1}, c_0^k) \Big)^2 \left( \frac{n}{\log \log n} \right)  \nonumber
        \\ & = \Bigg\{ \frac{1}{\sqrt{n \log \log n}} \sum_{i=1}^n \left[  \mathbb{I}_{ \{ Y_{i-k}^{i-1}=b_0^{k-1},Z_{i-k}^i=c_0^k   \} }  - P_{\Bar{Y}_{-k}^{-1},\Bar{Z}_{-k}^0} (b_0^{k-1}, c_0^k) \right]  \Bigg\} ^2, \label{eq_27}
\end{align}
is also bounded a.s. As a result, each summand in \eqref{eq:for_bounded} 
is a.s.\ bounded and, therefore, the entire expression tends a.s.\ to zero as $n \to \infty$, as claimed.

Exactly the same argument shows that, after being multiplied by $\sqrt{n}$, the other three relative entropies in \eqref{eq_27} also tend to zero, a.s., as $n \to \infty$, so that, a.s.,
\begin{align}
        \sqrt{n} \Big[ & \hat{I}_n^{(k)}  (\Xp \to \Yp\| \Zp) - I(\Bar{Y}_0 ; \Bar{X}_{-k}^0| \Bar{Y}_{-k}^{-1}, \Bar{Z}_{-k}^0 )  \Big] \nonumber  
        \\ & =  \frac{1}{\sqrt{n}} \sum_{i=1}^n \Bigg[ \log \left( \frac{ P_{\Bar{X}_{-k}^0,\Bar{Y}_0|\Bar{Y}_{-k}^{-1},\Bar{Z}_{-k}^0} (X_{i-k}^i, Y_i | Y_{i-k}^{i-1}, Z_{i-k}^i)}{P_{\Bar{Y}_0|\Bar{Y}_{-k}^{-1},\Bar{Z}_{-k}^0} (Y_i|Y_{i-k}^{i-1}, Z_{i-k}^i) P_{\Bar{X}_{-k}^0|\Bar{Y}_{-k}^{-1},\Bar{Z}_{-k}^0} (X_{i-k}^i |Y_{i-k}^{i-1}, Z_{i-k}^i)} \right)  \nonumber 
        \\  & \hspace{3.2in} - I(\Bar{Y}_0 ; \Bar{X}_{-k}^0| \Bar{Y}_{-k}^{-1}, \Bar{Z}_{-k}^0 ) \Bigg] + o(1). \label{eq_28}
\end{align}
The result of the theorem now follows by an application of 
the central limit theorem~\cite[Section~I.16]{chung:book} 
to the above partial sums of a functional of the chain 
$\{W_n\}$. Finally, the fact that the variance $\sigma^2$
exists as the stated limit, follows from~\cite[Theorem~3, p.~97]{chung:book}, 
and its finiteness
is a direct consequence of the fact that the 
alphabets $A$, $B$ and $C$ are finite and the 
random variables $\{ \log( \frac{\cdots}{\cdots} )\}$ being summed 
in~\eqref{eq_28} are uniformly bounded. \qed

\subsection{Proof of Theorem~\ref{thmX}}
\label{proofthmX}

Recall the hypothesis testing setup 
of Section~\ref{sec_hypothesis_test}. In that notation, 
the class of all possible transition matrices $Q = Q_\theta$ is 
parametrized by an $m^k \ell^k t^k (m \ell t-1)$-dimensional vector,
\begin{equation*}
    \theta = ( \theta_{i_1, \ldots, i_k, j_1, \ldots, j_k, s_1, \ldots, s_k  , i',j',s' }   ) = ( \theta_{i_1^k, j_1^k, s_1^k, i',j',s' }   ),
\end{equation*}
where $1\leq i_1, \ldots, i_k \leq m $, $1 \leq j_1, \ldots, j_k \leq \ell $, $1\leq s_1, \ldots, s_k \leq t $, and $(i',j',s') \in A \times B \times C$, $(i',j',s') \neq (m,\ell,t)$. Note that we have used the same notation 
as before for strings of symbols, $i_1^k \in A^k$, $j_1^k \in B^k$ 
and $s_1^k \in C^k$. 
The parameters $\theta$ that correspond to transition probability 
matrices $Q_\theta$ with positive transitions are those
in the open subset of $\mathbb{R}^{m^k \ell^k t^k(m \ell t-1)}$
given by,
\begin{align*}
\Theta = \Bigg\{   \theta 
& 
	\in (0,1)^{m^k \ell^k t^k(m \ell t-1)} : \\ 
& 
	\sum_{\substack{(i',j',s') \in A \times B 
	\times C \\ (i',j',s') \neq (m,\ell,t)}} 
	\theta_{i_1^k, j_1^k, s_1^k, i',j',s' } 
	< 1, \text{for all  } i_1^k \in A^k, j_1^k \in B^k 
	\text{  and  } s_1^k \in C^k   \Bigg\}, 
\end{align*}
so that each $\theta$ corresponds to a matrix
$Q_\theta$ with:
\begin{equation}
\label{eq_29}
    Q_\theta ( i',j',s'| i_1^k, j_1^k, s_1^k )=\left\{
                \begin{array}{ll}
                  \theta_{i_1^k, j_1^k, s_1^k, i',j',s' },     \hspace{1.5in} & \text{if } (i',j',s') \neq (m,\ell,t),  \\
                  1 - \sum_{\substack{(u,v,w)\neq (m,\ell,t)}} \theta_{i_1^k, j_1^k, s_1^k, i',j',s' },    & \text{if } (i',j',s') = (m,\ell,t).
                \end{array}
              \right.
\end{equation}

The parameter space $\Phi$ corresponding to the null hypotheses 
can be defined as the product
$\Phi = \Gamma^{x,z} \times \Gamma^y$, where,
\begin{align*}
    \Gamma^{x,z} = \Bigg\{   \gamma^{x,z} & = ( \gamma^{x,z}_{i_1^k, j_1^k, s_1^k, i',s'} ) \in (0,1)^{m^k \ell^k t^k (mt-1)} : 
    \\ & \sum_{\substack{ 1 \leq i' \leq m \\ 1 \leq s' \leq t \\ ( i', s') \neq (m,t)}} \gamma^{x,z}_{i_1^k, j_1^k, s_1^k, i',s'} < 1, \text{for all  } i_1^k \in A^k, j_1^k \in B^k \text{  and  } s_1^k \in C^k   \Bigg\}, 
\end{align*}
and,
\begin{align*}
    \Gamma^y = \Bigg\{   \gamma^y & = ( \gamma^y_{ s_1^k, s^{k+1}=s', j_1^k, j'} ) \in (0,1)^{ \ell^k t^{k+1} (\ell-1)} : 
    \\ & \sum_{\substack{ 1 \leq j' \leq \ell-1 }} \gamma^y_{j_1^k, s_1^{k+1}, j'} < 1, \text{for all  }  j_1^k \in B^k \text{  and  } s_1^{k+1} \in C^{k+1}   \Bigg\}.
\end{align*}
The space $\Phi$ is an open subset of 
$\mathbb{R}^{\ell^k t^k [m^k(mt-1) + t(\ell-1)]}$,
which can be naturally embedded in $\Theta$ via the 
map $h: \Phi \to \Theta$, where each component 
of $h(\phi) = h(\gamma^{x,z}, \gamma^y)$ is,
\begin{equation}
\label{eq_30}
    h_{i_1^k, j_1^k, s_1^k, i',j',s' }(\phi) = \gamma^{x,z}_{i_1^k, j_1^k, s_1^k, i',s'} . \gamma^y_{ j_1^k, s_1^{k+1},  j'}, 
\end{equation}
for all with $(i',s') \neq (m,t)$ and $j' \neq \ell$, 
and extending naturally to the `edge; 
cases $((m,t),j')$, $((i',s'),\ell)$ and  $((m,t),\ell)$. 
As a result, the corresponding transition probability 
matrix $Q_{h(\phi)}$ takes the form:
\begin{align}
&
	Q_{h(\phi)}( i',j',s'| i_1^k, j_1^k, s_1^k ) 
	\nonumber\\
&=
	\left\{
                \begin{array}{ll}
                \gamma^{x,z}_{i_1^k, j_1^k, s_1^k, i',s'} 
		\gamma^y_{ j_1^k, s_1^{k+1},  j'},     
		& (i',s') \neq (m,t)l,\; 
		j' \neq \ell,
		\\
                \left( 1- \sum_{(u',w') \neq (m,t)} 
		\gamma^{x,z}_{i_1^k, j_1^k, s_1^k, u',w'} \right) 
		\gamma^y_{ j_1^k, s_1^{k+1},  j'},     
		& (i',s') = (m,t),\; j' \neq \ell, 
		\\
                \gamma^{x,z}_{i_1^k, j_1^k, s_1^k, i',s'} 
		\left( 1- \sum_{v' \neq \ell} \gamma^y_{ j_1^k, s_1^{k+1},v'}
		\right),     
		& (i',s') \neq (m,t),\; j' = \ell,
                \\
		\left( 1- \sum_{(u',w') \neq (m,t)} 
		\gamma^{x,z}_{i_1^k, j_1^k, s_1^k, u',w'} \right) 
		\Big( 1- \sum_{v' \neq \ell} \gamma^y_{ j_1^k, s_1^{k+1},v'}
		\Big),   
		& (i',s') = (m,t),\; j' = \ell.
                \end{array}
	\right.      
	\label{eq_Qhphi}
\end{align}

The $\chi^2$-convergence stated in the theorem is
a consequence of Proposition~\ref{prop_3.5} combined
with a general result of Billingsley~\cite{billingsley:markov}.
Specifically, from Proposition~\ref{prop_3.5} we know that
the quantity of interest,
$2 n \hat{I}_n^{(k)} (\Xp \to \Yp \| \Zp)$, is equal
to the log-likelihood ratio $\Delta_n$ defined in \eqref{eq_18},
and~\cite[Theorem~6.2]{billingsley:markov} states
that $\Delta_n$ (or, in the notation of~\cite{billingsley:markov}),
$\Bar{\chi}^2_{t,t+1}(\hat{\phi})$),
converges in distribution 
to a $\chi^2$ with $r-c$ degrees of freedom, 
where $r,c$ are the dimensionalities of the parameter
spaces $\Theta$ and $\Phi$, respectively.
Since $r=m^k \ell^k t^k(m \ell t-1)$ 
and $c= \ell^k t^k [m^k(mt-1) + t(\ell-1)]$, the
limiting distribution has
\begin{equation*}
    r-c= \ell^k t^{k+1} (m^{k+1}-1) (\ell-1) \quad \text{degrees of freedom,}
\end{equation*}
as claimed.

The only thing that remains is to verify the 
two assumptions
of~\cite[Theorem~6.2]{billingsley:markov}, given 
Condition~6.1 of~\cite[p.~33]{billingsley:markov}.
Namely, we need to check that
the matrix $Q_{h(\phi)}$ has continuous third order 
partial derivatives throughout $\Phi$, 
and that the matrix $L(\phi)$ defined below has rank $c$.
The validity of the first assumption is obvious from
the definition of $Q_{h(\phi)}$ in \eqref{eq_Qhphi}:
Clearly $Q_{h(\phi)}$
has continuous third order partial derivatives with 
respect to every component $\gamma^{x,z}_{i_1^k, j_1^k, s_1^k, i',s'}$  
and $\gamma^y_{ j_1^k, s_1^{k+1},  j'}$ of $\phi$.

For the second assumption, let $s=m \ell t$ denote the alphabet size 
of $\{(X_n,Y_n,Z_n)\}$ and consider the 
$s^{t+1} \times c 
= (m \ell t)^{k+1} \times \ell^k t^k [m^k(mt-1) + t(\ell-1)] $ 
matrix $L(\phi)$, with entries given by the partial derivatives of 
each element $Q_{h(\phi)} ( i',j',s'| i_1^k, j_1^k, s_1^k )$ 
of the matrix $Q_{h(\phi)}$ with respect to every component of 
$\phi = (\gamma^{x,z}, \gamma^y)$. Then the row of $L(\phi)$ corresponding to $( i_1^k, j_1^k, s_1^k, i', j', s' )$ consists of all
the `$x,z$-derivatives',
\begin{equation}
\label{eq_31}
    \frac{\partial Q_{h(\phi)} ( i',j',s'| i_1^k, j_1^k, s_1^k ) }{\partial \gamma^{x,z}_{u_1^k, v_1^k, w_1^k, u',w'}},
\end{equation}
followed by all the `$y$-derivatives',
\begin{equation}
\label{eq_32}
    \frac{\partial Q_{h(\phi)} ( i',j',s'| i_1^k, j_1^k, s_1^k ) }{\partial \gamma^y_{ v_1^k, w_1^{k+1},  v'}}.
\end{equation}
From the definition of $Q_{h(\phi)}$ in~\eqref{eq_Qhphi}, 
we see that the derivatives in~\eqref{eq_31} are equal 
to zero unless $i_1^k=u_1^k$,$j_1^k=v_1^k$ and $s_1^k=w_1^k$, 
in which case:
\begin{equation*}
       \frac{\partial Q_{h(\phi)} ( i',j',s'| i_1^k, j_1^k, s_1^k ) }{\partial \gamma^{x,z}_{u_1^k, v_1^k, w_1^k, u',w'}}=\left\{
                \begin{array}{ll}
                  \gamma^y_{ j_1^k, s_1^{k+1},  j'},     \hspace{0.7in} & \text{if } (u',w')=(i',s') \neq (m,t),\; j' \neq \ell , 
                  \\ - \gamma^y_{ j_1^k, s_1^{k+1},  j'},     & \text{if } (u',w')  \neq (m,t)= (i',s'),\; j' \neq \ell,  
                  \\   1- \sum_{v' \neq \ell} \gamma^y_{ j_1^k, s_1^{k+1},  v'} ,    & \text{if } (u',w') = (i',s') \neq (m,t), \; j' = \ell,
                  \\   \sum_{v' \neq \ell} \gamma^y_{ j_1^k, s_1^{k+1},v'}-1,     & \text{if } (u',w')  \neq (m,t) = (i',s').\;  j' = \ell,
                  \\ 0, & \text{otherwise.}
                \end{array}
              \right.
\end{equation*}
Similarly, the derivatives in \eqref{eq_32} are equal to zero unless $j_1^k=v_1^k$ and $s_1^{k=1}=w_1^{k+1}$, in which case:
\begin{equation*}
    \frac{\partial Q_{h(\phi)} ( i',j',s'| i_1^k, j_1^k, s_1^k ) }{\partial \gamma^y_{ v_1^k, w_1^{k+1},  v'}}=\left\{
                \begin{array}{ll}
                  \gamma^{x,z}_{i_1^k, j_1^k, s_1^k, i',s'} ,     \hspace{0.7in} & \text{if } (i',s') \neq (m,t),\;  v'=j' \neq \ell , 
                  \\ 1- \sum_{(u',w') \neq (m,t)} \gamma^{x,z}_{i_1^k, j_1^k, s_1^k, u',w'} ,     & \text{if } (i',s') = (m,t),\; v'=j' \neq \ell,  
                  \\  -\gamma^{x,z}_{i_1^k, j_1^k, s_1^k, i',s'},     & \text{if } (i',s') \neq (m,t),\; v' \neq j' = \ell,
                  \\   \sum_{(u',w') \neq (m,t)} \gamma^{x,z}_{i_1^k, j_1^k, s_1^k, u',w'} -1,     & \text{if } (i',s') = (m,t),\; v' \neq j' = \ell,
                  \\ 0, & \text{otherwise.}
                \end{array}
              \right.
\end{equation*}
A straightforward (albeit tedious) 
examination of the above expressions shows that,
since all the components of all the parameters 
$\gamma^{x,z}$ and $\gamma^y$ are strictly positive, 
the matrix $L(\phi)$ has rank $c = \ell^k t^k [m^k(mt-1) + t(\ell-1)]$ 
for all $\phi \in \Phi$. This establishes the assumptions
of~\cite[Condition~6.1]{billingsley:markov} and completes the 
proof of the theorem.
\qed

\subsection{Proof of Corollary \ref{cor_3.4}}
Since $\{(X_n, Y_n, Z_n)\}$ 
is assumed to be an irreducible and aperiodic Markov chain of order $k$, 
the $(k+1)$-dimensional 
chain $\{  W_n = ( X_{n-k}^n, Y_{n-k}^n, Z_{n-k}^n   ) \}$ 
is also ergodic.

First we consider the a.s.-convergence in~\eqref{eq_13}.
Note that the computation leading to~\eqref{eq_27} earlier 
shows that, as $n \to \infty$,
\begin{equation*}
  \left( \sqrt{\frac{n}{\log \log n}}   \right)  
	D \Big(  \hat{P}_{Y_{-k}^{-1},Z_{-k}^0,n} 
	\Big\| P_{\Bar{Y}_{-k}^{-1},\Bar{Z}_{-k}^0} \Big)  
	= O \left( \sqrt{\frac{\log \log n}{n}}   \right) = o(1), 
	\quad \text{a.s.}
\end{equation*}
Similarly, an analogous bound holds for each each of the 
other three relative entropies in~\eqref{eq_26}
and, using these,~\eqref{eq_28} yields that, as $n\to\infty$:
\begin{align*}
        & \sqrt{\frac{n}{\log \log n}} \Big[ \hat{I}_n^{(k)}  (\Xp \to \Yp\| \Zp) - I(\Bar{Y}_0 ; \Bar{X}_{-k}^0| \Bar{Y}_{-k}^{-1}, \Bar{Z}_{-k}^0 )  \Big]   
        \\ &=  \frac{1}{\sqrt{n \log \log n}} 
	\sum_{i=1}^n \Bigg\{ \log \Bigg( \frac{ P_{\Bar{X}_{-k}^0,\Bar{Y}_0|\Bar{Y}_{-k}^{-1},\Bar{Z}_{-k}^0} (X_{i-k}^i, Y_i | Y_{i-k}^{i-1}, Z_{i-k}^i)}{P_{\Bar{Y}_0|\Bar{Y}_{-k}^{-1},\Bar{Z}_{-k}^0} (Y_i|Y_{i-k}^{i-1}, Z_{i-k}^i) P_{\Bar{X}_{-k}^0|\Bar{Y}_{-k}^{-1},\Bar{Z}_{-k}^0} (X_{i-k}^i |Y_{i-k}^{i-1}, Z_{i-k}^i)} \Bigg)     
        \\ & \hspace{4.1in} - I(\Bar{Y}_0 ; \Bar{X}_{-k}^0| \Bar{Y}_{-k}^{-1}, \Bar{Z}_{-k}^0 ) \Bigg\} + o(1).
\end{align*}
If $\sigma^2 > 0$, then the law of the iterated logarithm~\cite{chung:book} 
implies that the expression above is bounded, which gives the 
required result in~\eqref{eq_13}.
If $\sigma^2=0$, i
the same conclusion holds 
by~\cite[Theorem~17.5.4]{meyn-tweedie:book2}.

Next we consider the $L^1$ convergence rate in \eqref{eq_14}.
First we claim that each of the four relative entropies in~\eqref{eq_26}
converge to zero at a rate $O(1/n)$ in $L^1$. 
For the first one, 
using a first order Taylor expansion for the logarithm in 
the definition of the relative entropy gives,
\begin{align*}
&
	D\big(\hat{P}_{Y_{-k}^{-1},Z_{-k}^0,n} \big\| 
	P_{\Bar{Y}_{-k}^{-1},\Bar{Z}_{-k}^0} \big) 
	\\
&= 
	\sum_{b_0^{k-1}, c_0^k }  
	\hat{P}_{Y_{-k}^{-1},Z_{-k}^0,n} (b_0^{k-1}, c_0^k) 
	\left( \frac{\hat{P}_{Y_{-k}^{-1},Z_{-k}^0,n} (b_0^{k-1}, c_0^k)  
	-  P_{\Bar{Y}_{-k}^{-1},\Bar{Z}_{-k}^0} (b_0^{k-1}, c_0^k)}
	{P_{\Bar{Y}_{-k}^{-1},\Bar{Z}_{-k}^0} (b_0^{k-1}, c_0^k)} \right) 
	\left( \frac{1}{\zeta_n (b_0^{k-1}, c_0^k)}  \right),  
\end{align*}
where, for those $(b_0^{k-1}, c_0^k)$ for which the
first probability in the sum above is nonzero,
$\zeta_n (b_0^{k-1}, c_0^k)$ is a (possibly random) 
constant between 1 and,
$$\frac{\hat{P}_{Y_{-k}^{-1},Z_{-k}^0,n}(b_0^{k-1},c_0^k)}
	{P_{\Bar{Y}_{-k}^{-1},\Bar{Z}_{-k}^0} (b_0^{k-1}, c_0^k)},$$
while for the remaining $(b_0^{k-1}, c_0^k)$ 
it can be taken to be arbitrary.

Now define the random variable $S_n(b_0^{k-1}, c_0^k)$
as the difference
$$S_n(b_0^{k-1}, c_0^k)=  \hat{P}_{Y_{-k}^{-1},Z_{-k}^0,n} (b_0^{k-1}, c_0^k)  
-  P_{\Bar{Y}_{-k}^{-1},\Bar{Z}_{-k}^0} (b_0^{k-1}, c_0^k),$$
and let,
$$\rho_n(b_0^{k-1}, c_0^k)= [ \zeta_n(b_0^{k-1}, c_0^k) - 1 ]  
\frac{P_{\Bar{Y}_{-k}^{-1},\Bar{Z}_{-k}^0} (b_0^{k-1}, c_0^k)}
{S_n(b_0^{k-1}, c_0^k)}.$$ 
Then some simple algebra gives,
\begin{align*}
    D \Big(  \hat{P}_{Y_{-k}^{-1},Z_{-k}^0,n} & \| P_{\Bar{Y}_{-k}^{-1},\Bar{Z}_{-k}^0} \Big) 
    \\ & = \sum_{b_0^{k-1}, c_0^k }  S_n (b_0^{k-1}, c_0^k) \left(   \frac{1+ S_n (b_0^{k-1}, c_0^k)/ P_{\Bar{Y}_{-k}^{-1},\Bar{Z}_{-k}^0} (b_0^{k-1}, c_0^k) }{ 1+  \rho_n (b_0^{k-1}, c_0^k) S_n (b_0^{k-1}, c_0^k)/ P_{\Bar{Y}_{-k}^{-1},\Bar{Z}_{-k}^0} (b_0^{k-1}, c_0^k)  }      \right),
\end{align*}
where we note that each $\rho_n(b_0^{k-1}, c_0^k) \in [0,1]$. 
Moreover, using the following elementary inequality,
$(1+x)/(1+ \rho x) \leq 1 + x(1-\rho)$,
for $x>-1$, $\rho \in [0,1]$, we obtain
the bound,
\begin{align*}
    D  \big(  \hat{P}_{Y_{-k}^{-1},Z_{-k}^0,n} \big \| 
	P_{\Bar{Y}_{-k}^{-1},\Bar{Z}_{-k}^0} \big) 
	& \leq  \sum_{b_0^{k-1}, c_0^k }  S_n (b_0^{k-1}, c_0^k) \left( 1 + [ 1 - \rho_n (b_0^{k-1}, c_0^k) ] \frac{S_n (b_0^{k-1}, c_0^k)}{P_{\Bar{Y}_{-k}^{-1},\Bar{Z}_{-k}^0} (b_0^{k-1}, c_0^k)}  \right)
    \\ &  \leq  \sum_{b_0^{k-1}, c_0^k }  S_n (b_0^{k-1}, c_0^k) + \sum_{b_0^{k-1}, c_0^k } [ 1 - \rho_n (b_0^{k-1}, c_0^k) ]   \frac{S_n (b_0^{k-1}, c_0^k)^2}{P_{\Bar{Y}_{-k}^{-1},\Bar{Z}_{-k}^0} (b_0^{k-1}, c_0^k)}   
    \\ & \leq \sum_{b_0^{k-1}, c_0^k } \frac{S_n (b_0^{k-1}, c_0^k)^2}{P_{\Bar{Y}_{-k}^{-1},\Bar{Z}_{-k}^0} (b_0^{k-1}, c_0^k)},
\end{align*}
where we used 
the fact 
$\rho_n\in[0,1]$ and 
that each $S_n$ sums to zero by definition.

To show that the above relative entropy converges 
to zero in $L^1$ at rate $O(1/n)$, it suffices to prove
that each term in the last sum does. 
And for that it suffices to show that, for each $(b_0^{k-1}, c_0^k)$,
we have:
\begin{equation}
    \label{eq_33}
    n \BBE[ S_n(b_0^{k-1}, c_0^k)^2 ] = O(1), \quad \text{as } n \to \infty.
\end{equation}
But, as in~\eqref{eq_27}, $S_n(b_0^{k-1}, c_0^k)$ are simply the 
centered and normalized partial sums of a functional of the 
ergodic chain $\{W_n\}$,
\begin{equation*}
    S_n(b_0^{k-1}, c_0^k) = \frac{1}{n} \sum_{i=1}^n \left[       \mathbb{I}_{ \{ Y_{i-k}^{i-1}=b_0^{k-1},Z_{i-k}^i=c_0^k   \} }  - P_{\Bar{Y}_{-k}^{-1},\Bar{Z}_{-k}^0} (b_0^{k-1}, c_0^k)    \right],
\end{equation*}
so classical Markov chain theory tells us that
$n \BBE[ S_n(b_0^{k-1}, c_0^k)^2 ]$ converges to a finite
constant as $n \to \infty$; cf.~\cite[Theorem~3, p.~97]{chung:book}.
This implies \eqref{eq_33} and shows that,
\begin{equation*}
    \BBE \left|     D  \big(  \hat{P}_{Y_{-k}^{-1},Z_{-k}^0,n} \big \| 
	P_{\Bar{Y}_{-k}^{-1},\Bar{Z}_{-k}^0} \big)    \right| 
	= O\left(\frac{1}{n}\right), \quad \text{as }  n \to \infty.
\end{equation*}
And arguing exactly the same way shows that
the same result holds for the other three relative 
entropies in \eqref{eq_26}. 

Therefore, in order to prove the required $L^1$ bound in~\eqref{eq_14}, 
it suffices to show that,
\begin{align}
\BBE \Bigg| \frac{1}{n} \sum_{i=1}^n \left[A_i -  I(\Bar{Y}_0 ; \Bar{X}_{-k}^0| \Bar{Y}_{-k}^{-1}, \Bar{Z}_{-k}^0 ) \right]  \Bigg| = O\left(  \frac{1}{\sqrt{n}}  \right),  \label{eq_34}   
\end{align}
where,
$$A_i = \log \left( \frac{ P_{\Bar{X}_{-k}^0,\Bar{Y}_0|\Bar{Y}_{-k}^{-1},\Bar{Z}_{-k}^0} (X_{i-k}^i, Y_i | Y_{i-k}^{i-1}, Z_{i-k}^i)}{P_{\Bar{Y}_0|\Bar{Y}_{-k}^{-1},\Bar{Z}_{-k}^0} (Y_i|Y_{i-k}^{i-1}, Z_{i-k}^i) P_{\Bar{X}_{-k}^0|\Bar{Y}_{-k}^{-1},\Bar{Z}_{-k}^0} (X_{i-k}^i |Y_{i-k}^{i-1}, Z_{i-k}^i)} \right).$$
By the Cauchy-Schwarz inequality, we have,
\begin{align}
\sqrt{n} \BBE \Bigg| \frac{1}{n} \sum_{i=1}^n \left[A_i -  I(\Bar{Y}_0 ; \Bar{X}_{-k}^0| \Bar{Y}_{-k}^{-1}, \Bar{Z}_{-k}^0 ) \right]  \Bigg| & \leq \sqrt{ \frac{1}{n}  \BBE      \left|   \sum_{i=1}^n \left[A_i -  I(\Bar{Y}_0 ; \Bar{X}_{-k}^0| \Bar{Y}_{-k}^{-1}, \Bar{Z}_{-k}^0 ) \right]  \right| ^2    },  \label{Holder_ineq}
\end{align}
where, once again,~\cite[Theorem~3, p.~97]{chung:book} 
implies that the term insider the square root in~(\ref{Holder_ineq})
converges to a finite constant.
This establishes~\eqref{eq_34}, and completes the proof.
\qed

\subsection{Proof of Proposition \ref{prop_3.5}}
\label{appx_proof_prop_3.5}

Recall the expression from the definition of the log-likelihood
in~\eqref{eq_17} and the definition of $\Delta_n$ in~\eqref{eq_18}.
The first maximum in the definition of $\Delta_n$ is,
\begin{align*}
&
	\max_{\theta \in \Theta} 
	L_n(X_{-k+1}^n,Y_{-k+1}^n,Z_{-k+1}^n;\theta)\\
&= 
	\max_Q 
	\sum_{i=1}^n  \log \left( 
	Q_\theta(X_i,Y_i,Z_i | X_{i-k}^{i-1},Y_{i-k}^{i-1},Z_{i-k}^{i-1} ) 
	\right)\\
&=
	n 
	\frac{1}{n}\sum_{i=1}^n  
	\sum_{a_0^k, b_0^k, c_0^k } 
	\mathbb{I}_{ \{ X_{i-k}^i=a_0^k,Y_{i-k}^i=b_0^k,Z_{i-k}^i=c_0^k \} }  
	\log \left( 
	Q_\theta(a_k,b_k,c_k | a_0^{k-1},b_0^{k-1},c_0^{k-1} ) 
	\right),
\end{align*}
where the maximization over $Q$ is over all transition 
matrices $Q$ with all positive entries. Using the definition of the
empirical distribution we easily find that,
\begin{align*}
    \max_{\theta \in \Theta} 
	& \; L_n   (X_{-k+1}^n,Y_{-k+1}^n,Z_{-k+1}^n;\theta)
    \\ &= n \max_Q 
	\sum_{a_0^k, b_0^k, c_0^k } 
	 \hat{P}_{X_{-k}^0,Y_{-k}^0,Z_{-k}^0,n} (a_0^k, b_0^k, c_0^k)      \log \left( Q(a_k,b_k,c_k | a_0^{k-1},b_0^{k-1},c_0^{k-1} )  \right)
    \\ &= -n \min \Bigg\{     D\left( \hat{P}_{X_{-k}^0,Y_{-k}^0,Z_{-k}^0,n} (a_0^k, b_0^k, c_0^k) \Big| \Big| Q  \odot \hat{P}_{X_{-k}^{-1},Y_{-k}^{-1},Z_{-k}^{-1},n}  \right)
    \\ & \qquad \qquad \quad \; - \sum_{a_0^k, b_0^k, c_0^k } \hat{P}_{X_{-k}^0,Y_{-k}^0,Z_{-k}^0,n} (a_0^k, b_0^k, c_0^k)      \log \left( \frac{\hat{P}_{X_{-k}^0,Y_{-k}^0,Z_{-k}^0,n} (a_0^k, b_0^k, c_0^k)}{\hat{P}_{X_{-k}^{-1},Y_{-k}^{-1},Z_{-k}^{-1},n} (a_0^{k-1}, b_0^{k-1}, c_0^{k-1})}  \right) \Bigg\},
\end{align*}
where we used the notation,
\begin{equation*}
     (Q  \odot \hat{P}_{X_{-k}^{-1},Y_{-k}^{-1},Z_{-k}^{-1},n}) (a_0^k, b_0^k, c_0^k) = \hat{P}_{X_{-k}^{-1},Y_{-k}^{-1},Z_{-k}^{-1},n} (a_0^{k-1}, b_0^{k-1}, c_0^{k-1}) Q(a_k,b_k,c_k|a_0^{k-1}, b_0^{k-1}, c_0^{k-1}), 
\end{equation*}
for all $a_0^k \in A^{k+1}$, $b_0^k \in B^{k+1}$ and $c_0^k \in C^{k+1}$. 
The above minimum is obviously achieved by making the relative entropy equal to zero, that is, by taking,
\begin{equation*}
    Q(a_k,b_k,c_k|a_0^{k-1}, b_0^{k-1}, c_0^{k-1}) = \hat{P}_{X_0,Y_0,Z_0 | X_{-k}^{-1},Y_{-k}^{-1},Z_{-k}^{-1},n} (a_k,b_k,c_k | a_0^{k-1}, b_0^{k-1}, c_0^{k-1}),
\end{equation*}
so that,
\begin{equation}
\label{eq_35}
   \max_{\theta \in \Theta} L_n(X_{-k+1}^n,Y_{-k+1}^n,Z_{-k+1}^n;\theta) 
= n [ H ( \hat{X}_{-k}^{-1}, \hat{Y}_{-k}^{-1}, \hat{Z}_{-k}^{-1})
 - H ( \hat{X}_{-k}^0, \hat{Y}_{-k}^0, \hat{Z}_{-k}^0  ) ], 
\end{equation}
where $( \hat{X}_{-k}^0, \hat{Y}_{-k}^0, \hat{Z}_{-k}^0)  \sim \hat{P}_{X_{-k}^0,Y_{-k}^0,Z_{-k}^0,n} $.

For the second maximum in the definition of $\Delta_n$,
an analogous computation can be performed,
although it is slightly more complex as in this case 
it involves to two different maximizations. 
But since both of these are very similar to 
the one just computed, we only give an outline 
of the steps involved without providing all 
the details. 

Recall the the decomposition of $Q_\theta$ in~\eqref{eq_16} under the null. 
The second maximum in~\eqref{eq_17} can again be written in terms of
the empirical distribution,
\begin{align}
    \max_{\phi \in \Phi}&\,L_n(X_{-k+1}^n,Y_{-k+1}^n,Z_{-k+1}^n;h(\phi)) \nonumber
    \\ &= \max_{Q^{x,z},Q^y} \sum_{i=1}^n \log \Big( Q^{x,z}(X_i,Z_i|X_{i-k}^{i-1},Y_{i-k}^{i-1},Z_{i-k}^{i-1}) Q^y (Y_i|Y_{i-k}^{i-1},Z_{i-k}^i) \Big) \nonumber
    \\ &= \max_{Q^{x,z}} \sum_{i=1}^n \log \Big( Q^{x,z}(X_i,Z_i|X_{i-k}^{i-1},Y_{i-k}^{i-1},Z_{i-k}^{i-1}) \Big) + \max_{Q^y} \sum_{i=1}^n \log \Big( Q^y (Y_i|Y_{i-k}^{i-1},Z_{i-k}^i) \Big)       \nonumber
    \\ &= \max_{Q^{x,z}} \sum_{a_0^k, b_0^{k-1}, c_0^k } n \hat{P}_{X_{-k}^0,Y_{-k}^{-1},Z_{-k}^0,n} (a_0^k, b_0^{k-1}, c_0^k) \log \Big( Q^{x,z}(a_k,c_k|a_0^{k-1},b_0^{k-1},c_0^{k-1}) \Big)  \nonumber
    \\ & \qquad + \max_{Q^y} \sum_{b_0^k, c_0^k } n \hat{P}_{Y_{-k}^0,Z_{-k}^0,n} (b_0^k, c_0^k) \log \Big( Q^y (b_k|b_0^{k-1},c_0^k) \Big),\nonumber
\end{align}
and this can be further expressed in terms of entropies as,
\begin{align}
\max_{\phi \in \Phi}&\,L_n(X_{-k+1}^n,Y_{-k+1}^n,Z_{-k+1}^n;h(\phi)) \nonumber\\
     &= n \sum_{a_0^k, b_0^{k-1}, c_0^k }  \hat{P}_{X_{-k}^0,Y_{-k}^{-1},Z_{-k}^0,n} (a_0^k, b_0^{k-1}, c_0^k) \log \left( \frac{\hat{P}_{X_{-k}^0,Y_{-k}^{-1},Z_{-k}^0,n} (a_0^k, b_0^{k-1}, c_0^k)}{\hat{P}_{X_{-k}^{-1},Y_{-k}^{-1},Z_{-k}^{-1},n} (a_0^{k-1}, b_0^{k-1}, c_0^{k-1})} \right)  \nonumber
    \\ & \qquad +n \sum_{b_0^k, c_0^k }  \hat{P}_{Y_{-k}^0,Z_{-k}^0,n} (b_0^k, c_0^k) \log \left( \frac{\hat{P}_{Y_{-k}^0,Z_{-k}^0,n} (b_0^k, c_0^k)}{\hat{P}_{Y_{-k}^{-1},Z_{-k}^0,n} (b_0^{k-1}, c_0^k)} \right)    \nonumber
    \\&= n [ -H ( \hat{X}_{-k}^0, \hat{Y}_{-k}^{-1}, \hat{Z}_{-k}^0  ) 
	+ H ( \hat{X}_{-k}^{-1}, \hat{Y}_{-k}^{-1}, \hat{Z}_{-k}^{-1} )  
	-H (\hat{Y}_{-k}^0, \hat{Z}_{-k}^0 ) 
	+H (\hat{Y}_{-k}^{-1}, \hat{Z}_{-k}^0) ].
    \label{eq_36}
\end{align}

Finally, combining the results of~\eqref{eq_35} and~\eqref{eq_36} 
and using the chain rule for the entropy, we obtain
the required identity,
\begin{align*}
    \Delta_n &= 2 \left\{     \max_{\theta \in \Theta} L_n(X_{-k+1}^n,Y_{-k+1}^n,Z_{-k+1}^n;\theta) - \max_{\phi \in \Phi} L_n(X_{-k+1}^n,Y_{-k+1}^n,Z_{-k+1}^n;h(\phi))   \right\}
    \\ &= 2n [- H ( \hat{X}_{-k}^0, \hat{Y}_{-k}^0, \hat{Z}_{-k}^0 ) 
	+ H (  \hat{X}_{-k}^0, \hat{Y}_{-k}^{-1}, \hat{Z}_{-k}^0 )   
	+ H (  \hat{Y}_{-k}^0, \hat{Z}_{-k}^0 ) 
	- H (  \hat{Y}_{-k}^{-1}, \hat{Z}_{-k}^0  ) ]
    \\ &= 2n [-H(\hat{Y}_0|\hat{X}_{-k}^0,\hat{Y}_{-k}^{-1},\hat{Z}_{-k}^0) 
	+ H (\hat{Y}_0 |  \hat{Y}_{-k}^{-1}, \hat{Z}_{-k}^0)]
    \\&=2n \; \hat{I}_n^{(k)} (\Xp \to \Yp \| \Zp),
\end{align*}
by the definition of $\hat{I}_n^{(k)} (\Xp \to \Yp \| \Zp)$. \qed

\newpage

\bibliographystyle{plain}

\bibliography{ik.bib}

\def\cprime{$'$}
\begin{thebibliography}{10}

\bibitem{amblard:12}
P-O. Amblard and O.J.J. Michel.
\newblock The relation between {G}ranger causality and directed information
  theory: A review.
\newblock {\em Entropy}, 15(1):113, 2012.

\bibitem{anderson:11}
J.C. Anderson, H.~Kennedy, and K.A.C. Martin.
\newblock Pathways of attention: {Synaptic} relationships of frontal eye field
  to {V4}, lateral intraparietal cortex, and area 46 in macaque monkey.
\newblock {\em J. Neurosci.}, 31(30):10872--10881, July 2011.

\bibitem{anderson:14}
R.P. Anderson and M.~Porfiri.
\newblock Assessing significance of information flow in high dimensional
  dynamical systems with few data.
\newblock In {\em ASME 2014 Dynamic Systems and Control Conference}, pages
  V002T24A002--V002T24A002. American Society of Mechanical Engineers, 2014.

\bibitem{bachman:19}
P.~Bachman, R.D. Hjelm, and W.~Buchwalter.
\newblock Learning representations by maximizing mutual information across
  views.
\newblock In H.~Wallach, H.~Larochelle, A.~Beygelzimer, F.~d'Alch\'{e} Buc,
  E.~Fox, and R.~Garnett, editors, {\em Advances in Neural Information
  Processing Systems}, volume~32, Vancouver, BC, December 2019.

\bibitem{barnett:09}
L.~Barnett, A.B. Barrett, and A.K. Seth.
\newblock Granger causality and transfer entropy are equivalent for {Gaussian}
  variables.
\newblock {\em Phys. Rev. Lett.}, 103(23):238701, 2009.

\bibitem{barnett:12}
L.~Barnett and T.~Bossomaier.
\newblock Transfer entropy as a log-likelihood ratio.
\newblock {\em Phys. Rev. Lett.}, 109(13):138105, 2012.

\bibitem{battiti:94}
R.~Battiti.
\newblock Using mutual information for selecting features in supervised neural
  net learning.
\newblock {\em IEEE Trans. Neural Netw.}, 5(4):537--550, July 1994.

\bibitem{belghazi:18}
M.I. Belghazi, A.~Baratin, S.~Rajeshwar, S.~Ozair, Y.~Bengio, A.~Courville, and
  D.~Hjelm.
\newblock Mutual information neural estimation.
\newblock In {\em Proceedings of the 35th International Conference on Machine
  Learning}, ICML '18, pages 531--540, 2018.

\bibitem{billingsley:markov}
P.~Billingsley.
\newblock {\em Statistical inference for Markov processes}.
\newblock Statistical Research Monographs, Vol. II. The University of Chicago
  Press, Chicago, IL, 1961.

\bibitem{bruss:98}
D.~Bru{\ss}.
\newblock Optimal eavesdropping in quantum cryptography with six states.
\newblock {\em Phys. Rev. Lett.}, 81(14):3018--3012, October 1998.

\bibitem{cabuz:18}
S.~Cabuz and G.~Abreu.
\newblock Causal inference for multivariate stochastic process prediction.
\newblock {\em Inf. Sci.}, 448:134--148, June 2018.

\bibitem{chung:book}
K.L. Chung.
\newblock {\em Markov chains with stationary transition probabilities}.
\newblock Springer-Verlag, New York, NY, 1967.

\bibitem{cover:book2}
T.M. Cover and J.A. Thomas.
\newblock {\em Elements of information theory}.
\newblock John Wiley \& Sons, New York, NY, second edition, 2012.

\bibitem{csiszar:book2}
I.~Csisz{\'{a}}r and J.~K{\"{o}}rner.
\newblock {\em Information theory: Coding theorems for discrete memoryless
  systems}.
\newblock Cambridge University Press, Cambridge, U.K., second edition, 2011.

\bibitem{decampos:06}
L.M. De~Campos and N.~Friedman.
\newblock A scoring function for learning {Bayesian} networks based on mutual
  information and conditional independence tests.
\newblock {\em J. of Mach. Learn. Res.}, 7(10):2149--2187, 2006.

\bibitem{faes:13}
L.~Faes, G.~Nollo, and A.~Porta.
\newblock Compensated transfer entropy as a tool for reliably estimating
  information transfer in physiological time series.
\newblock {\em Entropy}, 15(1):198--219, 2013.

\bibitem{gallager:book}
R.G. Gallager.
\newblock {\em Information theory and reliable communication}.
\newblock John Wiley \& Sons, New York, NY, 1968.

\bibitem{gao:17}
W.~Gao, W.~Cui, and W.~Ye.
\newblock Directed information graphs for the {Granger} causality of
  multivariate time series.
\newblock {\em Physica A: Statistical Mechanics and its Applications},
  486:701--710, November 2017.

\bibitem{geweke:84}
J.F. Geweke.
\newblock Measures of conditional linear dependence and feedback between time
  series.
\newblock {\em J. Amer. Statist. Assoc.}, 79(388):907--915, 1984.

\bibitem{gierlichs:08}
B.~Gierlichs, L.~Batina, P.~Tuyls, and B.~Preneel.
\newblock Mutual information analysis: {A} generic side-channel distinguisher.
\newblock In {\em Proceedings of the 10th International Workshop on
  Cryptographic Hardware and Embedded Systems}, CHES 2008, pages 426--442,
  Washington, DC, August 2008.

\bibitem{granger:69}
C.W.J. Granger.
\newblock Investigating causal relations by econometric models and
  cross-spectral methods.
\newblock {\em Econometrica}, 37(3):424--438, 1969.

\bibitem{granger:74}
C.W.J. Granger and P.~Newbold.
\newblock Spurious regressions in econometrics.
\newblock {\em J. Econometrics}, 2(2):111--120, 1974.

\bibitem{gray-90:book}
R.M. Gray.
\newblock {\em Entropy and information theory}.
\newblock Springer-Verlag, New York, NY, 1990.

\bibitem{GraKie:Asymptotically:1980}
R.M. Gray and J.C. Kieffer.
\newblock Asymptotically mean stationary measures.
\newblock {\em Ann. Probab.}, 8(5):962--973, October 1980.

\bibitem{greenland:99}
S.~Greenland, J.~Pearl, and J.M. Robins.
\newblock Confounding and collapsibility in causal inference.
\newblock {\em Statist. Sci.}, 14(1):29--46, 1999.

\bibitem{greenwood:95}
P.E. Greenwood and W.~Wefelmeyer.
\newblock Efficiency of empirical estimators for {Markov} chains.
\newblock {\em Ann. Stat.}, 23(1):132--143, February 1995.

\bibitem{gregoriou:12}
G.G. Gregoriou, S.J. Gotts, and R.~Desimone.
\newblock Cell-type-specific synchronization of neural activity in {FEF} with
  {V4} during attention.
\newblock {\em Neuron}, 73(3):581--594, 2012.

\bibitem{gregoriou:09}
G.G. Gregoriou, S.J. Gotts, H.~Zhou, and R.~Desimone.
\newblock High-frequency, long-range coupling between prefrontal and visual
  cortex during attention.
\newblock {\em Science}, 324(5931):1207--1210, 2009.

\bibitem{gunduz-erkip:07}
D.~G\"{u}nd\"{u}z and E.~Erkip.
\newblock Lossless transmission of correlated sources over a multiple access
  channel with side information.
\newblock In {\em 2007 Data Compression Conference}, pages 83--92, March 2007.

\bibitem{haruna:13}
T.~Haruna and K.~Nakajima.
\newblock Permutation complexity and coupling measures in hidden {Markov}
  models.
\newblock {\em Entropy}, 15(9):3910--3930, September 2013.

\bibitem{haskovec:15}
J.~Haskovec.
\newblock A note on the consensus finding problem in communication networks
  with switching topologies.
\newblock {\em Applicable Analysis}, 94(5):991--998, 2015.

\bibitem{weissman-et-al:di}
J.~Jiao, H.H. Permuter, L.~Zhao, Y.-H. Kim, and T.~Weissman.
\newblock Universal estimation of directed information.
\newblock {\em IEEE Trans. Inform. Theory}, 59(10):6220--6242, October 2013.

\bibitem{kim:08}
Y-H. Kim.
\newblock A coding theorem for a class of stationary channels with feedback.
\newblock {\em IEEE Trans. Inform. Theory}, 54(4):1488--1499, April 2008.

\bibitem{kinney:14}
J.B. Kinney and G.S. Atwal.
\newblock Equitability, mutual information, and the maximal information
  coefficient.
\newblock {\em Proc. Natl. Acad. Sci. U.S.A.}, 111(9):3354--3359, 2014.

\bibitem{kirchgassner:book}
G.~Kirchg{\"a}ssner, J.~Wolters, and U.~Hassler.
\newblock {\em Introduction to modern time series analysis}.
\newblock Springer, Berlin, 2012.

\bibitem{ctw-isit:21}
I.~Kontoyiannis, L.~Mertzanis, A.~Panotonoulou, I.~Papageorgiou, and
  M.~Skoularidou.
\newblock Revisiting context-tree weighting for {Bayesian} inference.
\newblock In {\em 2021 IEEE International Symposium on Information Theory
  (ISIT)}, pages 2906--2911, Melbourne, Australia, July 2021.

\bibitem{BCT-JRSSB:22}
I.~Kontoyiannis, L.~Mertzanis, A.~Panotonoulou, I.~Papageorgiou, and
  M.~Skoularidou.
\newblock {Bayesian Context Trees}: {Modelling} and exact inference for
  discrete time series.
\newblock {\em J. R. Stat. Soc. Series B}, 84(4):1287--1323, September 2022.

\bibitem{skoularidou-K:16}
I.~Kontoyiannis and M.~Skoularidou.
\newblock Estimating the directed information and testing for causality.
\newblock {\em IEEE Trans. Inform. Theory}, 62(11):6053--6067, November 2016.

\bibitem{kramer:phd}
G.~Kramer.
\newblock {\em Directed information for channels with feedback}.
\newblock PhD thesis, Swiss Federal Institute of Technology, Zurich,
  Switzerland, 1998.

\bibitem{kramer:14}
G.~Kramer.
\newblock Information networks with in-block memory.
\newblock {\em IEEE Trans. Inform. Theory}, 60(4):2105--2120, January 2014.

\bibitem{kugiumtzis:13}
D.~Kugiumtzis.
\newblock Partial transfer entropy on rank vectors.
\newblock {\em Eur. Phys. J. Spec. Top.}, 222(2):401--420, 2013.

\bibitem{lehmann:book}
E.L. Lehmann and G.~Casella.
\newblock {\em Theory of point estimation}.
\newblock Springer, New York, NY, 2006.

\bibitem{lungu-pap-K:22}
V.~Lungu, I.~Papageorgiou, and I.~Kontoyiannis.
\newblock Bayesian change-point detection via context-tree weighting.
\newblock In {\em 2022 IEEE Workshop on Information Theory (ITW)}, Mumbai,
  India, November 2022.

\bibitem{lungu-arxiv:22}
V.~Lungu, I.~Papageorgiou, and I.~Kontoyiannis.
\newblock Change-point detection and segmentation of discrete data using
  {Bayesian} context trees.
\newblock {\em arXiv e-prints}, \texttt{2203.04341 [stat.ME]}, March 2022.

\bibitem{mackay:book}
D.J.C. MacKay.
\newblock {\em Information theory, inference and learning algorithms}.
\newblock Cambridge University Press, New York, NY, 2003.
\newblock Available at \texttt{www.inference.org.uk/itprnn}.

\bibitem{marko:73}
H.~Marko.
\newblock The bidirectional communication theory: {A} generalization of
  information theory.
\newblock {\em IEEE Trans. Commun.}, 21(1):1335--1351, December 1973.

\bibitem{massey:90}
J.L. Massey.
\newblock Rate-distortion in near-linear time.
\newblock In {\em 1990 International Symposium on Information Theory and its
  Applications (ISITA)}, pages 303--305, November 1990.

\bibitem{meyn-tweedie:book2}
S.P. Meyn and R.L. Tweedie.
\newblock {\em {Markov chains and stochastic stability}}.
\newblock Cambridge University Press, London, U.K., second edition, 2009.
\newblock Published in the Cambridge Mathematical Library. 1993 edition online:
  {\tt probability.ca/MT/}.

\bibitem{mijatovic:21}
G.~Mijatovic, Y.~Antonacci, T.~Loncar-Turukalo, L.~Minati, and L.~Faes.
\newblock An information-theoretic framework to measure the dynamic interaction
  between neural spike trains.
\newblock {\em IEEE Trans. Biomed. Eng.}, 68(12):3471--3481, 2021.

\bibitem{olive:04}
D.J. Olive.
\newblock Does the {MLE} maximize the likelihood?
\newblock Unpublished manuscript, available at {\tt
  parker.ad.siu.edu/Olive/simle.pdf}, August 2004.

\bibitem{weissman-et-al:11}
H.H. Permuter, Y-H. Kim, and T.~Weissman.
\newblock Interpretations of directed information in portfolio theory, data
  compression, and hypothesis testing.
\newblock {\em IEEE Trans. Inform. Theory}, 57(6):3248--3259, June 2011.

\bibitem{permuter-et-al:09}
H.H. Permuter, T.~Weissman, and A.J. Goldsmith.
\newblock Finite state channels with time-invariant deterministic feedback.
\newblock {\em IEEE Trans. Inform. Theory}, 55(2):644--662, February 2009.

\bibitem{neurspik}
C.J. Quinn, T.P. Coleman, N.~Kiyavash, and N.G. Hatsopoulos.
\newblock Estimating the directed information to infer causal relationships in
  ensemble neural spike train recordings.
\newblock {\em J. Comput. Neurosci.}, 30(1):17--44, December 2011.

\bibitem{quinn:15}
C.J. Quinn, N~Kiyavash, and T.P. Coleman.
\newblock Directed information graphs.
\newblock {\em IEEE Trans. Inform. Theory}, 61(12):6887--6909, December 2015.

\bibitem{rahimian:13}
M.A. Rahimian and V.M. Preciado.
\newblock Detection and isolation of link failures under the agreement
  protocol.
\newblock In {\em Decision and Control (CDC), 2013 IEEE 52nd Annual Conference
  on}, pages 7364--7369, December 2013.

\bibitem{schamberg:20}
G.~Schamberg and T.P. Coleman.
\newblock Measuring sample path causal influences with relative entropy.
\newblock {\em IEEE Trans. Inform. Theory}, 66(5):2777--2798, May 2020.

\bibitem{schreiber:00}
T.~Schreiber.
\newblock Measuring information transfer.
\newblock {\em Phys. Rev. Lett.}, 85(2):461, 2000.

\bibitem{shibuya:11}
T.~Shibuya, T.~Harada, and Y.~Kuniyoshi.
\newblock Reliable index for measuring information flow.
\newblock {\em Phys. Rev. E}, 84(6):061109, December 2011.

\bibitem{shojaie:22}
A.~Shojaie and E.B. Fox.
\newblock Granger causality: {A} review and recent advances.
\newblock {\em Annual Review of Statistics and Its Application}, 9:289--319,
  March 2022.

\bibitem{spinney:17}
R.E. Spinney, M.~Prokopenko, and J.T. Lizier.
\newblock Transfer entropy in continuous time, with applications to jump and
  neural spiking processes.
\newblock {\em Phys. Rev. E}, 95(3):032319, 2017.

\bibitem{steuer:02}
R.~Steuer, J.~Kurths, C.O. Daub, J.~Weise, and J.~Selbig.
\newblock The mutual information: {Detecting} and evaluating dependencies
  between variables.
\newblock {\em Bioinformatics}, 18(suppl\_2):S231--S240, October 2002.

\bibitem{tishby:99}
N.~Tishby, F.~Pereira, and W.~Bialek.
\newblock The information bottleneck method.
\newblock In {\em Proceedings of the 37th Annual Allerton Conference on
  Communication and Computation}, pages 368--377, Allerton, IL, 1999.

\bibitem{vergara:14}
J.R. Vergara and P.A. Est{\'e}vez.
\newblock A review of feature selection methods based on mutual information.
\newblock {\em Neural Comput. and Applic.}, 24:175--186, 2014.

\bibitem{weissman-et-al:13}
T.~Weissman, Y-H. Kim, and H.H. Permuter.
\newblock Directed information, causal estimation, and communication in
  continuous time.
\newblock {\em IEEE Trans. Inform. Theory}, 59(3):1271--1287, March 2013.

\end{thebibliography}

\end{document}